\documentclass[twoside,twocolumn,9pt]{article}
\usepackage{extsizes}%
\usepackage[super,sort&compress,comma]{natbib}%
\usepackage[version=3]{mhchem}%
\usepackage[left=1.5cm, right=1.5cm, top=1.785cm, bottom=2.0cm]{geometry}%
\usepackage{balance}%
\usepackage{times,mathptmx}%
\usepackage{sectsty}%
\usepackage{graphicx}%
\usepackage{lastpage}%
\usepackage[format=plain,justification=justified,singlelinecheck=false,font={stretch=1.125,small,sf},labelfont=bf,labelsep=space]{caption}%
\usepackage{float}%
\usepackage{fancyhdr}%
\usepackage{fnpos}%
\usepackage[english]{babel}%
\addto{\captionsenglish}{%
   }%
\usepackage{array}%
\usepackage{droidsans}%
\usepackage{charter}%
\usepackage[T1]{fontenc}%
\usepackage[usenames,dvipsnames]{xcolor}%
\usepackage{setspace}%
\usepackage[compact]{titlesec}%
\usepackage{hyperref}%
\usepackage{cleveref}%
\usepackage{siunitx}%
\usepackage{soul}%
\usepackage{mathtools}%

\usepackage{epstopdf}

\definecolor{cream}{RGB}{222,217,201}

\graphicspath{{./figs/}}%

\newcommand{\FD}{\textrm{f}_\textrm{d}}

\begin{document}

\pagestyle{fancy}
\thispagestyle{plain}
\fancypagestyle{plain}{

\fancyhead[C]{\includegraphics[width=18.5cm]{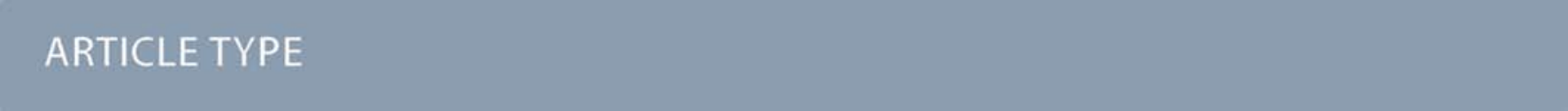}}
\fancyhead[L]{\hspace{0cm}\vspace{1.5cm}\includegraphics[height=30pt]{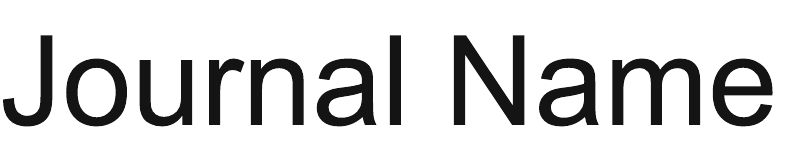}}
\fancyhead[R]{\hspace{0cm}\vspace{1.7cm}\includegraphics[height=55pt]{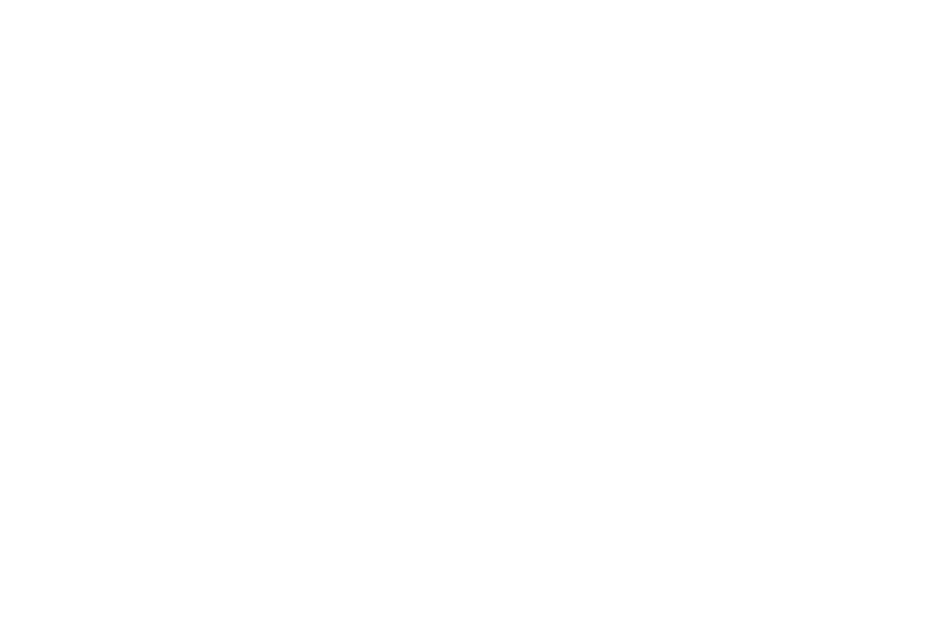}}
\renewcommand{\headrulewidth}{0pt}
}

\makeFNbottom
\makeatletter
\renewcommand\LARGE{\@setfontsize\LARGE{15pt}{17}}
\renewcommand\Large{\@setfontsize\Large{12pt}{14}}
\renewcommand\large{\@setfontsize\large{10pt}{12}}
\renewcommand\footnotesize{\@setfontsize\footnotesize{7pt}{10}}
\makeatother

\renewcommand{\thefootnote}{\fnsymbol{footnote}}
\renewcommand\footnoterule{\vspace*{1pt}%
\color{cream}\hrule width 3.5in height 0.4pt \color{black}\vspace*{5pt}} 
\setcounter{secnumdepth}{5}

\makeatletter 
\renewcommand\@biblabel[1]{#1}            
\renewcommand\@makefntext[1]%
{\noindent\makebox[0pt][r]{\@thefnmark\,}#1}
\makeatother 
\renewcommand{\figurename}{\small{Fig.}~}
\sectionfont{\sffamily\Large}
\subsectionfont{\normalsize}
\subsubsectionfont{\bf}
\setstretch{1.125} 
\setlength{\skip\footins}{0.8cm}
\setlength{\footnotesep}{0.25cm}
\setlength{\jot}{10pt}
\titlespacing*{\section}{0pt}{4pt}{4pt}
\titlespacing*{\subsection}{0pt}{15pt}{1pt}

\fancyfoot{}
\fancyfoot[LO,RE]{\vspace{-7.1pt}\includegraphics[height=9pt]{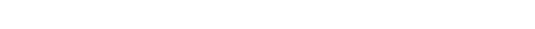}}
\fancyfoot[CO]{\vspace{-7.1pt}\hspace{13.2cm}\includegraphics{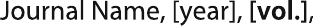}}
\fancyfoot[CE]{\vspace{-7.2pt}\hspace{-14.2cm}\includegraphics{head_foot/RF}}
\fancyfoot[RO]{\footnotesize{\sffamily{1--\pageref{LastPage} ~\textbar  \hspace{2pt}\thepage}}}
\fancyfoot[LE]{\footnotesize{\sffamily{\thepage~\textbar\hspace{3.45cm} 1--\pageref{LastPage}}}}
\fancyhead{}
\renewcommand{\headrulewidth}{0pt} 
\renewcommand{\footrulewidth}{0pt}
\setlength{\arrayrulewidth}{1pt}
\setlength{\columnsep}{6.5mm}
\setlength\bibsep{1pt}

\makeatletter 
\newlength{\figrulesep} 
\setlength{\figrulesep}{0.5\textfloatsep} 

\newcommand{\topfigrule}{\vspace*{-1pt}%
\noindent{\color{cream}\rule[-\figrulesep]{\columnwidth}{1.5pt}} }

\newcommand{\botfigrule}{\vspace*{-2pt}%
\noindent{\color{cream}\rule[\figrulesep]{\columnwidth}{1.5pt}} }

\newcommand{\dblfigrule}{\vspace*{-1pt}%
\noindent{\color{cream}\rule[-\figrulesep]{\textwidth}{1.5pt}} }

\makeatother

\twocolumn[
  \begin{@twocolumnfalse}
\vspace{3cm}
\sffamily
\begin{tabular}{m{4.5cm} p{13.5cm} }
  
  \includegraphics{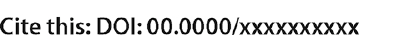} & \noindent\LARGE{\textbf{Tumbling with a Limp: Local Asymmetry in Water's Hydrogen Bond Network and its Consequences}} \\
  \vspace{0.3cm} & \vspace{0.3cm} \\
  
                                  & \noindent\large{Hossam Elgabarty$^{\ast}$\textit{$^{a}$} and Thomas D. K\"uhne\textit{$^{a,b}$}} \\

  \includegraphics{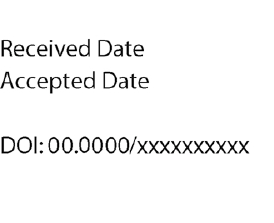} & \noindent\normalsize{Ab initio molecular dynamics simulations of liquid water under equilibrium ambient conditions, together with a novel energy decomposition analysis, have recently shown that a substantial fraction of water molecules exhibit a significant asymmetry between the strengths of the two donor and/or the two acceptor interactions. We refer to this recently unraveled aspect as the "local asymmetry in the hydrogen bond network". We discuss how this novel aspect was first revealed, and provide metrics that can be consistently employed on simulated water trajectories to quantify this local heterogeneity in the hydrogen bond network and its dynamics. We then discuss the static aspects of the asymmetry, pertaining to the frozen geometry of liquid water at any given instant of time and the distribution of hydrogen bond strengths therein, and also its dynamic characteristics pertaining to how fast this asymmetry decays and the kinds of molecular motions responsible for this decay. Following this we discuss the spectroscopic manifestations of this asymmetry, from ultrafast X-ray absorption spectra to infrared spectroscopy and down to the much slower terahertz regime. Finally, we discuss the implications of these findings in a broad context and its relation to the current notions about the structure and dynamics of liquid water.}\\

\end{tabular}

 \end{@twocolumnfalse} \vspace{0.6cm}

  ]

\renewcommand*\rmdefault{bch}\normalfont\upshape
\rmfamily
\section*{}
\vspace{-1cm}


\footnotetext{\textit{$^{a}$~Dynamics of Condensed Matter and Center for Sustainable Systems Design, Chair of Theoretical Chemistry, University of Paderborn, Warburger Str. 100, D-33098, Paderborn, Germany. E-mail: hossam.elgabarty@upb.de}}
\footnotetext{\textit{$^{b}$~Paderborn Center for Parallel Computing and Institute for Lightweight Design, University of Paderborn, Warburger Str. 100, D-33098, Paderborn, Germany.}}




\restoregeometry

\section{Introduction}

The many remarkable properties exhibited by liquid water remain subject of
intensive on going research. While the exceptional nature of many of these
properties warrant investigation in their own right, it is also clear that a
better understanding of the molecular basis of the properties of liquid water
is of paramount relevance to biology, chemistry, materials science, and
geology, just to name a few disciplines.\cite{Franks1972,Eisenberg2005} It is
currently recognized that water is no-longer merely the background against
which many reactions, and most biochemical reactions, are taking place. Water
is just as important as any of the key (bio)chemical players on any
stage.\cite{Gerstein1998,Ball2008} Shortcomings in our understanding of liquid
water are, \emph{ipso facto}, limitations to our understanding of the chemistry in aqueous or humid environments.
Despite of intensive efforts, many fundamental details are unknown and many
questions remain heavily debated. It is remarkable that we basically do not
understand exactly how microwaves heat water,\cite{Lunkenheimer2017} nor how
heat dispersion through the HB network actually works. Similarly, the
microscopic mechanism underlying water's dielectric function in the low
frequency region have been under constant debate, so is the basic tetrahedral
structure of water, which has been challenged based on some interpretations of
water's thermodynamic and spectroscopic characteristics.\cite{Tokushima2008}

At the heart of the properties of liquid water is its dynamic hydrogen bond
(HB) network with its fluctuating local tetrahedral geometry and 
collective behaviour. In fact, the remarkable properties of water and also
its so-called anomalies can ultimately be traced back in one way or another to
this dynamic network when combined with the small molecular size and molecular
polarity.\cite{Finney2004,Kaatze2018} Because of this, characterizing
individual HB dynamics and localized motional modes is far from sufficient
for a full understanding of the complexities of the underlying dynamical
processes. The collective and emergent nature of the ensuing properties simply
prohibits such a naive reductionist view. As a drastic example of this, it is
enough to recognize that while it is true that water without HBs would have
been a gas at ambient conditions, it is also true that without the HB
cooperativity, water would have been a very viscous liquid that would hardly
be similar to the matrix of life that we know. This is because the average HB
energy in liquid water is an order of magnitude higher than thermal
fluctuations at room temperature\cite{Khaliullin2013} and it is only through
the collective motion that water flows as it does. It is also because of these
cooperative modes that the dynamics of water remains a difficult problem for
analytical theory that cannot be adequately handled by molecular hydrodynamic
theory or continuum-model-based theories, which is why atomistic computer simulations
are usually required to interpret experimental findings. Indeed, simulations
have provided insights that were crucial for the interpretation of some of the
most important spectroscopic investigations of
water.\cite{Bakker2010,Soper2014,Perakis2016}


In this article, we are going to review some of our efforts in unraveling a particular
aspect of water's HB network, which is the local asymmetry in the strengths of
HBs donated or accepted from/to a water molecule in bulk liquid water under
ambient conditions.\cite{Kuehne2013,Khaliullin2013,Kuehne2014,Zhang2013,Zhang2015,Elgabarty2015,Ojha2018,Elgabarty2019} While it is well known that the HB strength
in water or in ice exhibit a broad distribution trivially due to thermal
fluctuations, the novel and non-trivial aspect that was revealed is that in
liquid water (but not in hexagonal ice), thermal motion leads to a significant
population of molecules, where the strength of the two donor or acceptor
interactions have an extreme disparity with as much as 25\% of the molecules
having one donor (acceptor) interaction that is more than six times stronger
than the other donor (acceptor) interaction, and a significant population
having both kinds of interactions highly asymmetric. This interesting feature
of water's HB network was revealed using \emph{ab initio} molecular dynamics (AIMD)
simulations \cite{Kuehne2007,Kuehne2014b} combined with a condensed phase energy decomposition analysis
based on absolutely localized molecular orbitals
(ALMO-EDA).\cite{Kuehne2013,Elgabarty2015} The strength of this method is that it
permits a rigorous quantification of the amount of charge-transfer by locating
the variationally lowest energy state without charge-transfer.\cite{Khaliullin2013}. Thereby, the
issues of under/overestimation of charge-transfer and contamination due to
charge overlap effects are circumvented.\cite{Khaliullin2007} The decomposition of the collective
interaction energy in the condensed phase into physically meaningful components
revealed this significant instantaneous asymmetry within the strength of the local
donor–acceptor contacts. 

In the following sections of this review we will discuss
how this novel aspect was first revealed, its static aspects pertaining to the
frozen geometry of liquid water at any given instant of time, and its dynamic
characteristics pertaining to how fast this asymmetry decays and the kinds of
molecular motions that brings about such decay. Following this we discuss the
consequences of this asymmetry, in particular how it shows up or influences
different kinds of spectroscopy from ultrafast X-ray absorption (XA) spectra to
line shapes in infrared (IR) spectroscopy and down to the much slower terahertz (THz)
regime. Finally, we discuss the implications of these findings in a broad
context and in particular, its relation to the current notions about the
structure and dynamics of liquid water.

\section{ALMO-EDA and the Discovery of the Instantaneous Asymmetry in Liquid Water's
  Hydrogen Bond Network}
  
\subsection{ALMO-EDA: Energy decomposition analysis based on absolutely
  localized molecular orbitals}

Intermolecular bonding is the end result of a complicated interplay of
multipolar electrostatic interactions, polarization effects, geometric
distortions, Pauli repulsion, and charge-transfer interactions (also known as
covalent, delocalization, or donor-acceptor orbital
interactions).\cite{Khaliullin2007} The goal of an electron-decomposition
analysis (EDA) is to isolate these physically relevant components from the
total energy of an interacting system composed of several subunits. In case of
liquid water the subunits are of course the individual water molecules, and
the total energy is, for example, that obtained from an \emph{ab initio}
calculation (wave function-based or density functional-based) for a system
composed of many interacting water molecules. In this case, what is of interest
to us is bulk liquid water as simulated using periodic boundary conditions.

The EDA method based on ALMOs, has
the key advantage that it permits the calculation of the energy lowering
associated with the electron transfer due to hydrogen bonding, using a
self-consistent approach that includes cooperativity effects, which are
essential for a correct description of the HB network. The ALMO-EDA method works by
separating the total interaction energy ($\Delta{E}_{TOT}$) into molecular
frozen density ($\Delta{E}_{FRZ}$), molecular polarization
($\Delta{E}_{POL}$), and pair-wise charge-transfer ($\Delta{E}_{CT}$)
terms, i.e.
\begin{equation}
  \label{eq:almo_eda}
  \Delta{E}_{TOT} = \Sigma_{i}^{\textrm{molecules}} \left (\Delta{E}_{FRZ}^i +
    \Delta{E}_{POL}^i \right ) + \Sigma_{i,j>i}^{\textrm{pairs}} \Delta{E}_{CT}^{i,j}.
\end{equation}
Therein, ($\Delta{E}_{FRZ}^i$) is the energy change in each molecule $i$ within the system,
which is brought about by bringing infinitely separated distorted molecules
(at their given geometries in bulk water) into the bulk without any relaxation
of the electronic structure of the molecules. This roughly parallels the idea
of atomization energy, except that the building blocks in this case are not
individual atoms, but molecules with all their internal degrees of freedom
(nuclear positions and electron density) totally frozen. Thus, ($\Delta{E}_{POL}^i$) is due to the intramolecular relaxation of each molecule's electrons in the
field of all other molecules in the system. These two terms are molecular and
additive, \emph{i.e.} the total contribution to the energy of the system is
simply a sum over all the molecular contributions. The ALMO method is designed
such that this relaxation constrains the molecular orbitals to remain strictly
localized on their respective molecules, hence the name absolutely localized
molecular orbitals.\cite{Khaliullin2007}
Finally, the contribution from parwise two-body charge-transfer interactions
($\Delta{E}_{CT}$) is computed as the energy difference between the relaxed
(\emph{i.e.} polarized) ALMOs and the fully optimized and delocalized
molecular orbitals, which are obtained by a full self-consistent field
calculation of the whole system. Hence, $\Delta{E}_{CT}^{i,j}$ accounts for
the energy lowering due to electron transfer from the occupied ALMOs on one
molecule $i$ to the virtual orbitals of another molecule $j$. In the current
implementation this term is computed using a single noniterative
Rayleigh-Schr\"odinger (RS) perturbative correction $\Delta{E}_{CT}^{\textrm{RS}}$
starting from the converged ALMO solution, i.e. neglecting a much smaller
second-order energy change due to induction that accompanies such
occupied-virtual mixing.  Detailed descriptions of the ALMO-EDA terms with
their mathematical details and algorithmic implementations have been given by
\citeauthor{Khaliullin2007}\cite{Khaliullin2013, Khaliullin2007,Khaliullin2008,Khaliullin2009,Kuehne2013, Elgabarty2015})

\begin{figure}
  \centering
  \includegraphics[width=0.7\columnwidth]{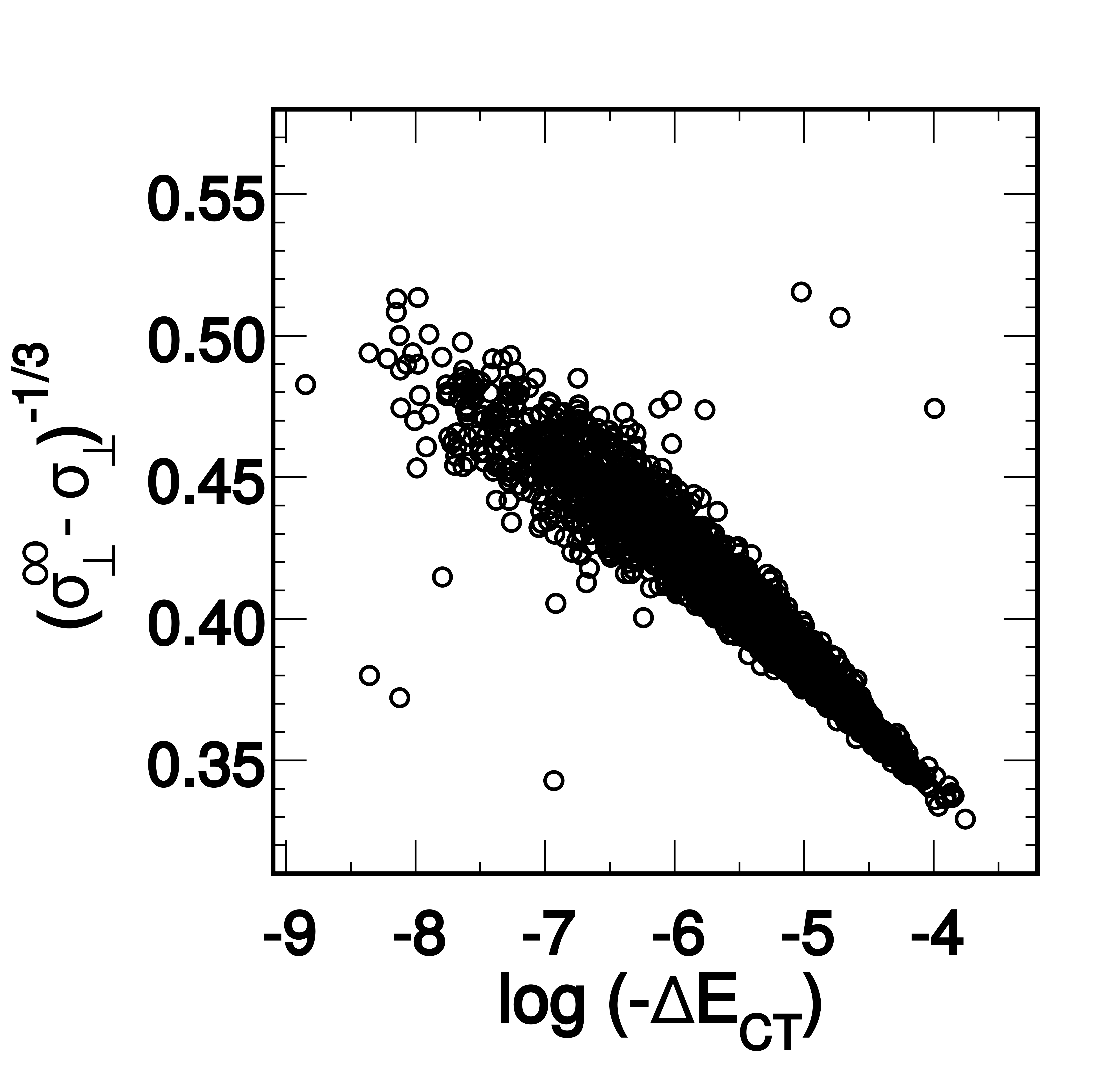}
  \caption{A derived linear model relating the orthogonal component of the $^1$H shielding tensor in water ($\sigma_{\perp}$) to the stabilization energy due to charge-transfer ($\Delta{E}_{CT})$). The shielding tensor component in the limit of a HB-free water molecule is denoted as $\sigma_{\perp}^\infty$. Reproduced from Ref.~\citenum{Elgabarty2015} / CC BY.}
  \label{fig:nmr}
\end{figure}

Most importantly for the present context is that the RS perturbative occupied-virtual electron transfer can be further decomposed into
forward and back-donation components for each pair $(i,j)$ of molecules:
\begin{equation}
  \Delta{E}_{CT}^{i,j} \approx \Delta{E}_{CT}^{i,j (\textrm{RS})} = \sum_{i,j>i}{ \Delta{E}_{i
      \rightarrow j}^{\textrm{RS}}
    + \Delta{E}_{j \rightarrow i}^{\textrm{RS}}}.
\end{equation}
Such a transfer of electrons from an acceptor to a donor is characteristic for
hydrogen bonding, where a HB acceptor mostly acts as an electron donor, and vice versa.\cite{Scheiner1997} The charge-transfer term is also, by
definition, directly related to the covalency of a hydrogen
bond.\cite{Khaliullin2007,Grabowski2011,Elgabarty2015} Moreover, we have also
shown that the charge-transfer term can be accurately estimated using proton
nuclear magnetic resonance spectroscopy (\cref{fig:nmr}), providing a means to
experimentally quantify this important aspect of intermolecular
bonding.\cite{Elgabarty2015}
The elegance of $\Delta{E}_{CT}$ (and also $\Delta{Q}_{CT}$, the fractional
charge-transfer that is responsible for the $\Delta{E}_{CT}$ term) is that
these metrics are solely based on the two-body delocalization-energy and charge-transfer components, which are readily computable for any given molecular
configuration, and as such can be directly used to study static HB networks, as well as the HB rearrangement dynamics without incorporating any (arbitrary) geometry-based criterion of a HB.


\subsection{Instantaneous asymmetry in the first coordination shell of liquid
  water}

\begin{figure}
  \centering
  \includegraphics[width=1.0\columnwidth]{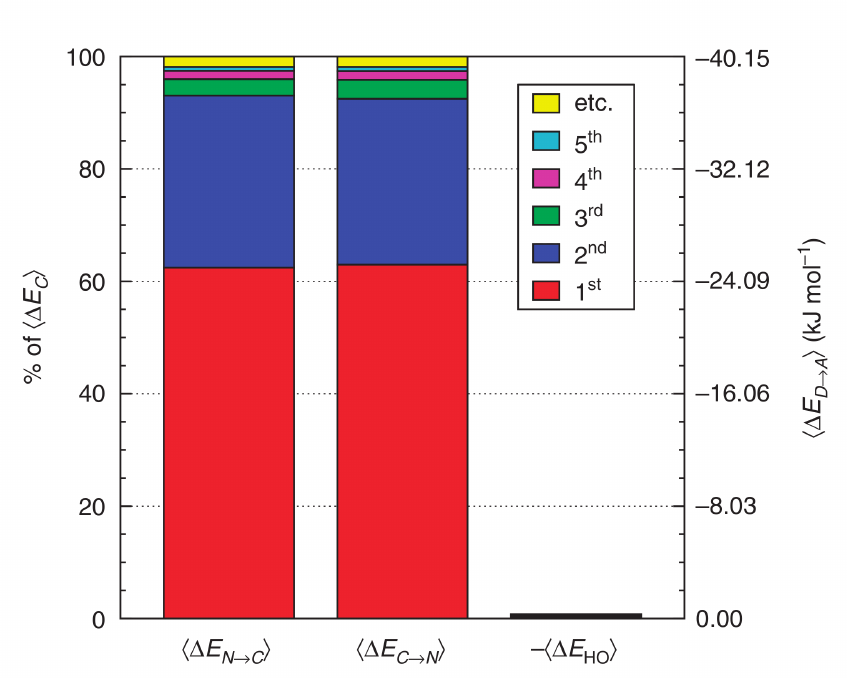}
  \caption{Average contributions of the five strongest accpetor ($\Delta{E}_{N\rightarrow{}C}$) and donor ($\Delta{E}_{C\rightarrow{}N}$) interactions. The right-most column shows that higher-order delocalization does not significantly contribute to the overall binding (See also shaded areas in \cref{fig:R_and_theta}). Angular brackets within the axes labels denote averaging over all central water molecules and all AIMD snapshots. Reproduced from Ref.~\citenum{Kuehne2013}.}
  \label{fig:almo1}
\end{figure}

When the ALMO-EDA technique was used to analyze snapshots extracted from
AIMD trajectories of liquid water, the
fractional electron transfer through a HB turned out to be a few
milli-electrons only, but nevertheless this contributes significantly to the
total HB energy and is roughly equal to the contribution of the polarization
term.\cite{Khaliullin2009,Khaliullin2013}
\Cref{fig:almo1} depicts the average of the strongest five acceptor and donor
contributions to the average delocalization energy of a molecule
($\Delta{E}_{CT}$). Examination of the depicted pattern of charge-transfer
interactions reveals that electron delocalization is dominated by two strong
donor (accpetor) interactions that together are responsible for 93\% of the
delocalization energy of a single molecule. The third and the fourth
strongest donor (acceptor) interactions contribute only 5\% and correspond to
back donation of electrons to (from) the remaining two first-shell neighbours. This is to say that a HB donor still weakly donates electrons to the unoccupied orbitals
of the HB acceptor, and \emph{vice versa}. The remaining 2\% correspond to
the delocalization energy to (from) the second and more distant coordination
shells.

\begin{figure}
  \centering \includegraphics[width=1.0\columnwidth]{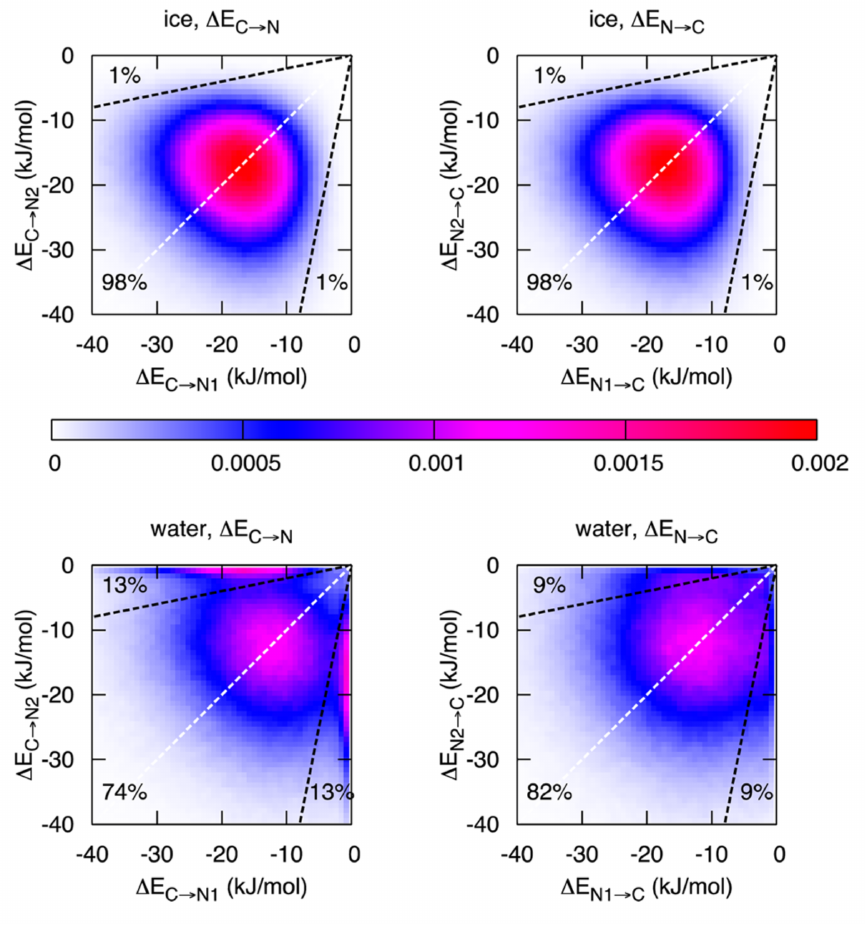}
  \caption{Distribution of molecules in ice and liquid water according to the
    strength of the first two strongest donor ($\Delta{E}_{C\rightarrow{N}}$)
    and acceptor ($\Delta{E}_{N\rightarrow{C}}$) interactions. The X or Y axis are 
assigned randomly, i.e. independently from the HB energies. The dashed
    white lines are the lines of the ideal symmetry $\gamma_D=0$ and
    $\gamma_A=0$. The dashed black lines correspond to $\gamma_D=0.8$ and
    $\gamma_A=0.8$, respectively. Reprinted with permission from Ref.~\citenum{Kuehne2014}. Copyright 2014 American Chemical Society.}
  \label{fig:ice_almo}
\end{figure}

Comparison of the strengths of the first and second strongest donor–acceptor
interactions ($\langle \Delta{E}_{CT}\rangle \sim 25 $ and $\sim 12$
\si{\kilo\joule\per\mole}, respectively) with that of the water dimer
($\sim\SI{9}{\kilo\joule\per\mole}$) immediately suggests that each water molecule can be considered to form on average two donor and two acceptor bonds, in agreement
with the usual tetrahedral structural picture of the coordination shell in
liquid water. What is striking in \cref{fig:almo1}, however, is the substantial
difference in the strengths of the first and second strongest interactions,
which implies that a large fraction of water molecules experience a
significant asymmetry in their local environment. The same pattern is also
clear when the energies of the two strongest donors or acceptors are plotted
together, which is depicted in the bottom two panels of
\cref{fig:ice_almo}. In a simple-minded picture, where the strength of a HB
directly correlates with its length, this instantaneous asymmetry would correspond to a
structural picture where, due to thermal agitation, many water molecules are
in distorted local tetrahedral environments. We will shortly scrutinize this
simple geometric picture in more quantitative rigor, and anticipating the
discussion, we assert that things are not that simple regarding the relation
between HB strength and the geometry in the first coordination shell.

\begin{figure}
  \centering
  \includegraphics[width=1.0\columnwidth]{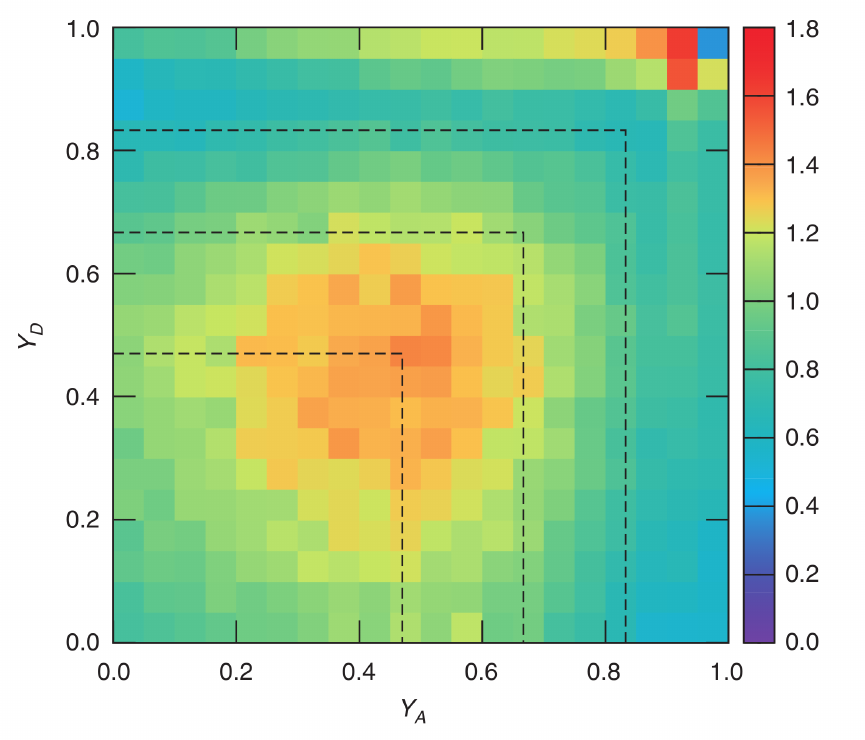}
  \caption{The normalized probability density function of the asymmetry
    parameters $\gamma_A$ and $\gamma_D$, respectively. The probability of finding a
    molecule in a bin can be found by dividing the corresponding density value
    by the number of bins (that is, 400). The dashed black lines at $\gamma
    \approx \frac{1}{2},\frac{2}{3},\frac{5}{6}$ partition all
    molecules into four groups of equal sizes. Reproduced from Ref.~\citenum{Kuehne2013}.}
  \label{fig:gamma_histogram}
\end{figure}

To characterize the aforementioned asymmetry of the two strongest donor or
acceptor contacts of a molecule, a dimensionless asymmetry parameter 
\begin{equation}
  \gamma_D = 1 - \frac{\Delta{E}_{C\rightarrow N^{2\textrm{nd}}}}{\Delta{E}_{C\rightarrow N^{1\textrm{st}}}} \hspace{2em} \gamma_A = 1 - \frac{\Delta{E}_{N\rightarrow C^{2\textrm{nd}}}}{\Delta{E}_{N\rightarrow C^{1\textrm{st}}}}
  \label{eq:gamma}
\end{equation}
was introduced, where $\gamma_A$ is used to denote the asymmetry within the acceptor and $\gamma_D$ for the donor interactions. The asymmetry parameters are obviously zero
when the two strongest contacts are equally strong (perfectly symmetric) and
equals to one when the second contact ceases to exist. The joint probability
distribution of the molecules according to their $\gamma$ asymmetry parameters
is depicted in \cref{fig:gamma_histogram} together with the lines indicating the
quartiles separating the molecules into four groups of equal sizes. The shape
of the distribution demonstrates that most molecules form highly asymmetric
donor or acceptor contacts at any instance of time. For example, the line at
$\gamma \approx \frac{1}{2}$ indicates that for 75\% of the molecules either
$\gamma_A$ or $\gamma_D$ is more than 0.5, which means that for the majority
of molecules the strongest donor or acceptor contact is at least two times
stronger than the second strongest.  Furthermore, the peak in the region of
high $\gamma$ in \cref{fig:gamma_histogram} indicates the presence of
molecules with significant asymmetry in HB strengths, both for donated and accepted
HBs simultaneously. This is also clear from the line at
$\gamma \approx \frac{5}{6}$, indicating that 25\% of the molecules have the
strongest interaction six times stronger than the second-strongest.

\begin{figure}
  \centering
  \includegraphics[width=0.3\columnwidth]{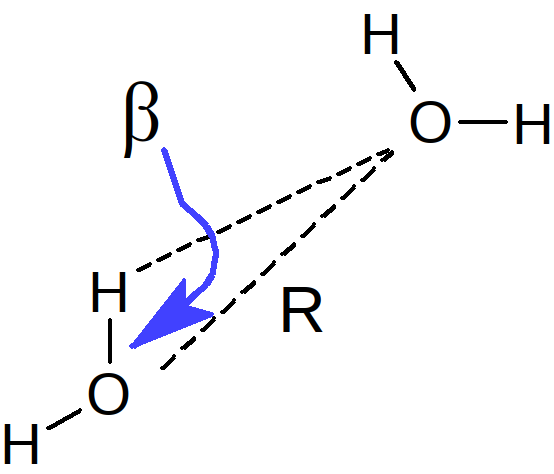}
  \caption{Definition of HB distance $R$ and angle $\beta$.}
  \label{fig:r_beta}
\end{figure}

\begin{figure}
  \centering
  \includegraphics[width=1.0\columnwidth]{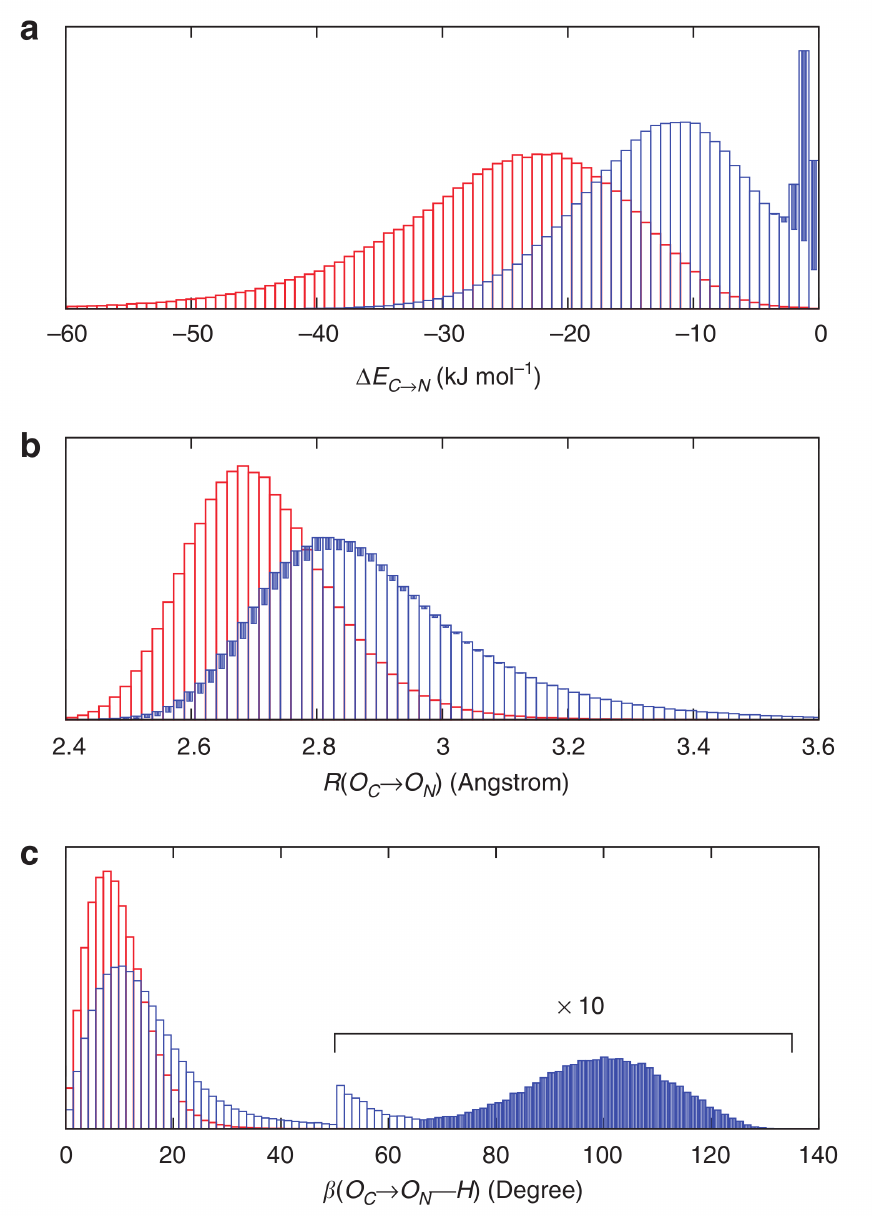}
  \caption{Energetic and geometric characteristics of the instantaneous
    asymmetry. The probability distribution of (a) HB strength, (b)
    intermolecular distance R and (c) HB angle $\beta$ for the first
    (red) and second (blue) strongest donor interactions C-N. Filled areas
    shows the contribution of configurations, for which back donation to a
    nearby donor is stronger than donation to the second strongest
    acceptor. Distributions for acceptor interactions N-C are almost
    identical and not shown. Reproduced from Ref.~\citenum{Kuehne2013}.}
  \label{fig:R_and_theta}
\end{figure}

To understand the origin of this asymmetry, the geometry of donor–acceptor
pairs involved in the first and second strongest interactions was compared. It
was found that the strength of the interaction is greatly affected by the intermolecular HB distance $R = d(O_D-O_A)$ and to a weaker extent by the
HB angle $\beta = \angle{(O_DO_AH)}$ (See \cref{fig:r_beta} for a depiction of $R$ and $\beta$), while the other geometric parameters
have only minor influence on $\Delta{E}_{CT}$.
The histograms of $\Delta{E}_{CT}$, $R$, and $\beta$ for the two strongest
donor interactions are shown in \cref{fig:R_and_theta}. The strong overlap
between all the histograms suggests that some second strongest interactions
have the same energetic and geometric characteristics as the strongest
contacts. This implies that the observed electronic asymmetry cannot be
attributed to the presence of two distinct types of HBs — weak and
strong! It is rather a result of continuous deformations of a typical
HB. Another important conclusion that can be made from the distributions in
\cref{fig:R_and_theta} is that relatively small variations of the HB distance
$\Delta{R}\sim\SI{0.2}{\angstrom}$ and angle
$\Delta{\theta}\sim\SIrange[range-phrase=\text{~to~}]{5}{10}{\degree}$ entails remarkable changes in the strength and electronic structure of HBs. Specifically, the average intermolecular oxygen--oxygen distances for
the strongest and second strongest interactions differ only by
$\Delta{R}\sim\SI{0.2}{\angstrom}$ and are not large enough to justify
asymmetric models of liquid water. This is a sober reminder that with the
highly non-linear relation between the HB strength and HB geometry,
simple-minded projections of the features of one onto the another can be very
misleading, and a high degree of asymmetry in the former does not imply a
dramatic distortion in the latter. Furthermore, analysis of the structure of
the molecular chains defined by the strongest bonds (that is, one donor
and one acceptor for each molecule) shows that their directions are random and
without any long-range order (\emph{e.g.} rings, spirals or zig-zags) on the
length scale of the simulation box ($\sim\SI{15}{\angstrom}$).

Because of the slow decay of the distribution tails in \cref{fig:R_and_theta},
a quantification of the concentration of single-donor and single-acceptor
molecules was not attempted. Defining such configurations using a distance,
angle or energy cutoff is an unavoidably arbitrary procedure. A quantitative
analysis of the network, which was performed for the same molecular
configurations than in \citeauthor{Kuehne2009} shows that according to the
commonly used geometric definitions of hydrogen
bonds,\cite{Wernet2004,Luzar1996,Kuo2004} the structure of water is distorted
tetrahedral with only a small fraction of broken bonds.

Finally, with an asymmetry in liquid water that is due to thermal distortions,
it is very interesting to compare this with the results of ALMO-EDA in
hexagonal ice, which is generally regarded to be a solid with symmetric HBs. Such a
comparison has been performed and indeed does provide further insight into the
origin of the asymmetry in liquid water and its relation to the HB network
structure.\cite{Kuehne2014} \Cref{fig:ice_almo} shows the joint distribution
of the strengths of the two pairs of donor and acceptor HBs in ice. The
ensuing two-dimensional distribution is characterized by the peak centered at
\SI{-18.7}{\kilo\joule\per\mole}, large deviation from the average values
$(\sigma = \SI{7.1}{\kilo\joule\per\mole})$, and a small correlation coefficient of
0.1. Such a tiny correlation coefficient indicates that the two HBs are
essentially independent from each other and the asymmetry in ice (points in
the distribution that are significantly displaced from the diagonal line) arises
trivially from the very broad distribution of HB strengths. As in the case
of liquid water, this asymmetry is a result of thermal fluctuations around the
average symmetric structure.
The distribution of the strength of the donor interactions in liquid water
exhibits a drastically different pattern with two pronounced features. In
addition to a broad peak resembling the one for ice, there is a sharp peak in
the region of high $\gamma_D$. The center of the first peak is shifted to
lower energies $(\SI{-12}{\kilo\joule\per\mole})$ and is somewhat broader than
that of ice. The second peak indicates the presence of molecules with one
intact and one broken donor HB
$(\Delta{E}_{C\rightarrow{N}} > \SI{-1}{\kilo\joule\per\mole})$. To estimate the
fraction of molecules responsible for the sharp asymmetric feature, we draw a
somewhat arbitrary boundary at $\gamma_D = 0.8$ (dashed line in
\Cref{fig:ice_almo}), which divides the distribution into the regions of
ice-like configurations and highly asymmetric
configurations. \Cref{fig:ice_almo} shows that 26\% (18\%) of molecules in
the liquid phase are characterized by $\gamma_D > 0.8$ ($\gamma_A > 0.8$)
compared to 2\% in ice.

In hexagonal ice, the uncorrelated thermal motion of molecules around their crystallographic sites
broadens the range of HB energies, and thus creates a noticeable asymmetry
in the donor and acceptor contacts of each water molecule. This is trivially a
broad distribution without any particular correlation between the strengths of
the pair of donor (or acceptor) interactions at any center. In water, while
the majority of molecules still exhibit HB patterns similar to those in ice
and retains a four-fold coordination with only moderately distorted
tetrahedral configurations, there is the drastic difference of a large
fraction of molecules with very weak HBs, which are elongated by as much as
0.5 Å and exhibits a wide range of angular distortions. A detailed analysis is given
in Ref. \citenum{Kuehne2014}. These results imply that the traditional view
of water as a four-fold coordinated nearly tetrahedral liquid is more appropriate
than the recently proposed asymmetric
model.\cite{Wernet2004,Odelius2006,Tokushima2008} However, the substantial
fraction of molecules with broken HBs undoubtedly affects the physical
properties and chemical behavior of liquid water. Starting from
\Cref{sec:electr-sign-asymm} we investigate some of the consequences of this
aspect of liquid water.

\subsection{Contrasting the asymmetry in donor versus acceptor interactions}

\Cref{fig:ice_almo} reveals an interesting difference between donor and
acceptor interactions in liquid water. While the corresponding plots in ice
are identical, they are rather different in liquid water. Thermal disorder
affects the strengths of donor and acceptors interactions in a quantitatively
different manner. First, while also significant, the fraction of molecules with broken acceptor bonds (18\% with $\gamma_D > 0.8$), is lower than the
corresponding fraction in $\gamma_A$ (26\%). Furthermore, the distribution of
the acceptor interactions does not exhibit a high-$\gamma_A$ peak, which would
match the high-$\gamma_D$ peak. This difference indicates that within the broken HBs,
only the donor of electrons remains under-coordinated, while the acceptor
(i.e., hydrogen atom) forms a HB with another donor that becomes
overcoordinated. The existence of a significant fraction of overcoordinated donors is supported
by the relatively large contribution of the third interaction shown in
\Cref{fig:almo1} and is consistent with the well-known fact that the
distribution of electron acceptors around a water molecule is more disordered
than that of the donors.\cite{Agmon2011} This phenomenon is attributed to the
existence of the so-called “negativity track” between the lone pairs of a
water molecule, which facilitates the disordered motion of electron acceptors
around the central donor.\cite{Becke1990,Keutsch2001} This asymmetry in donor versus acceptor interactions is also reminiscent of the asymmetry in solvation energy of anions and cations,\cite{Remsing2018} and a possible relation between both phenomena 

\subsection{Dynamics of the asymmetry}

\begin{figure}
  \centering
  \includegraphics[width=1.0\columnwidth]{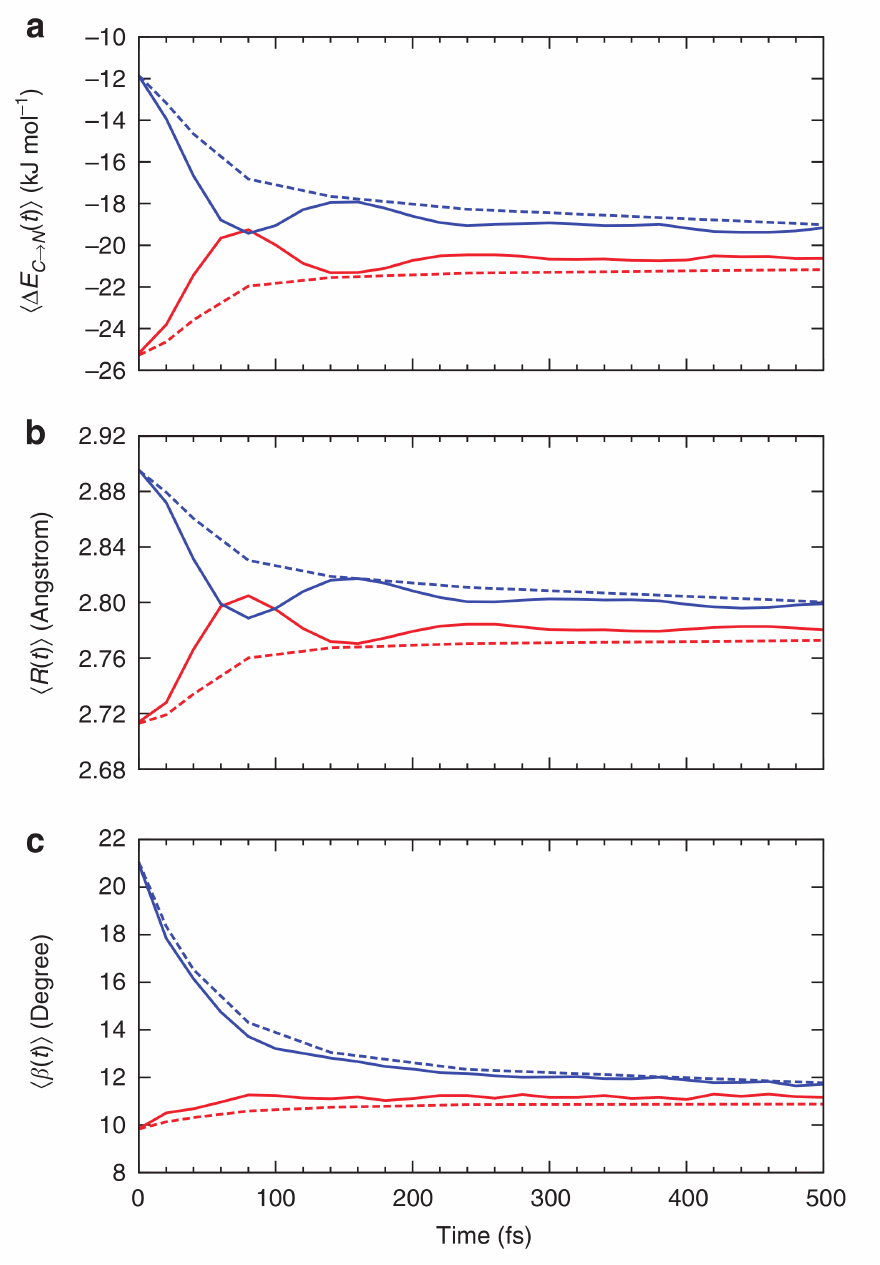}
  \caption{Relaxation of the instantaneous asymmetry. Time-dependence of the
    (a) average HB strength, (b) intermolecular distance R and (c) HB angle $\beta$ for the first
    $\Delta{E}_{C\rightarrow N^{1\textrm{st}}}$(red) and second
    $\Delta{E}_{C\rightarrow N^{2\textrm{nd}}}$ (blue) strongest donor
    interactions. Solid lines shows the instantaneous values, whereas the
    dashed lines correspond to the time-average values. Time-dependent
    characteristics of acceptor interactions $\Delta{E}_{N\rightarrow C}$
    are almost identical and not shown. Reproduced from Ref.~\citenum{Kuehne2013}.}
  \label{fig:gamma_dynamics}
\end{figure}

The overlapping distributions in \Cref{fig:R_and_theta} suggest that, despite
the difference in the strength of the donor–acceptor contacts, their nature is
similar and that the strongest interacting pair can become the second strongest in
the process of thermal motion and vice versa. To estimate the time scale of
this process, it is necessary to examine how the average energies of the first
two strongest interactions fluctuate in time. The instantaneous values at
time $t$ (solid lines in \cref{fig:gamma_dynamics}a) were calculated using the
ALMO EDA terms for 3501 snapshots separated by 20 fs (448128 local
configurations) by averaging over time origins $\tau$ separated by 100 fs and
over all surviving triples:
\begin{equation}
  \langle \Delta{E}_{C\rightarrow{N}}(t) \rangle = \frac{1}{T}
  \sum\limits_{\tau=1}^{T}\frac{1}{M(\tau,t)}\sum\limits_{C=1}^{M(\tau,t)}\Delta{E}_{C\rightarrow{N}}(\tau+t),
\end{equation}
where $M(\tau,t)$ is the number of triples that survived from time $\tau$ to
$\tau+t$. A triple is considered to survive a specified time interval if the
central molecule has the same two strongest-interacting neighbours in all
snapshots within this interval.

\Cref{fig:gamma_dynamics}a shows that the strength of the first two strongest
interactions oscillates rapidly and that $\sim\SI{80}{\femto\second}$ after an
arbitrarily chosen time origin, the strongest interaction becomes
slightly weaker than the second strongest (note that first and second refer to
their order at $t = 0$). The amplitude of the oscillations decreases and the
strengths of both interactions approaches the average value of
$\sim\SI{20}{\kilo\joule\per\mole}$ on the time scale of hundreds of
femtoseconds. The decay of the oscillations indicates fast decorrelation of
the time-separated instantaneous values because of the strong coupling of a
selected pair of molecules with its surroundings. In other words, while the
energy of a particular HB fluctuates around its average value, this bond has
approximately equal chances of becoming weak or strong after a certain period
of time independently of its strength at $t = 0$. This effect is due to the
noise introduced by the environment and can be observed in ultrafast IR
spectroscopy experiments.\cite{Fecko2003} The time averages shown in
\cref{fig:gamma_dynamics} are physically meaningful and can be calculated
accurately only for the time intervals that are shorter than the average
lifetime of a HB.\cite{Luzar2000} The small residual asymmetry that is still
present after \SI{500}{\femto\second} (\cref{fig:gamma_dynamics}a) is an
indication of the slow non-exponential relaxation behaviour that characterizes
the kinetics of many processes in liquid water.\cite{Luzar1996} Specifically,
the non-exponential relaxation of the HB lifetime is due to the coupling
between HB fluctuations and diffusion,\cite{Luzar1996} and the non-exponential
tail in \cref{fig:gamma_dynamics}a a likely manifestation of this process. At variance, the
fast relaxation component in \cref{fig:gamma_dynamics}a is more
correlated with HB stretch motions and closely matches the time scale of 170fs
found from the spectral diffusion of the OH stretch peak in IR
spectroscopy.\cite{Ramasesha2013,Fecko2003,Perakis2016}

In addition to the instantaneous values of $\Delta{E}_{C\rightarrow{N}}(t)$
\emph{at} time t, the dashed lines in \cref{fig:gamma_dynamics}a show the
corresponding averages \emph{over} time interval of length $t$. These values
were calculated by averaging over time origins $\tau$, all snapshots lying in
the time interval from $\tau$ to $\tau + t$ and over all surviving triples:
\begin{equation}
  \overline{\langle \Delta{E}_{C\rightarrow{N}}(t) \rangle} = \frac{1}{T}\sum\limits_{\tau=1}^{T}\frac{1}{t+1}
  \sum\limits_{\kappa=0}^{t}\frac{1}{M(\tau,t)}\sum\limits_{C=1}^{M(\tau,t)}\Delta{E}_{C\rightarrow{N}}(\tau+\kappa).
\end{equation}
The dashed lines in \cref{fig:gamma_dynamics}a show that any neighbour-induced
asymmetry in the electronic structure of a water molecule can be observed only
with an experimental probe with a time-resolution of tens of femtoseconds or
less. On longer time scales, the asymmetry is destroyed by the thermal motion
of molecules and only the average symmetric structures can be observed in
experiments with low temporal resolution. An examination of the time
dependence of all two-body and some three-body geometric parameters that
characterize the relative motion of molecules reveals the mechanism of the
relaxation. The curves in \cref{fig:gamma_dynamics}a and b show that the
asymmetry is almost completely lost on the time scale of a single cycle of the
HB stretch motion ($\sim\SI{200}{\per\centi\meter}$, which corresponds to
$\sim\SI{170}{\femto\second}$). \hl{It is worth noting here that the decay of the time correlation function of the instantaneous fluctuations in energy $\langle \delta{E}_1(0)\,\delta{E}_2(t)$ follows the same time scale}. Relaxation of the asymmetry is thus primarily caused by low-frequency vibrations of the molecules relative to each other. The minor differences in the behaviour of the curves, in particular at 80 fs, indicate that the relaxation of the asymmetry is possibly also
influenced by some faster degrees of freedom. The temporal changes in the HB
angles towards the average value (\cref{fig:gamma_dynamics}c) show that
librations of molecules play a minor role in the relaxation process.

The kinetics and mechanism of the asymmetry relaxation presented here are
supported by data from ultrafast IR spectroscopy, which can directly
observe the spectral diffusion of the OH stretch peak with a fast component
that monoexponentially decays at 170
fs.\cite{Fecko2005,Ramasesha2013,Perakis2016} In conclusion, it is important
to point out that the dynamics of the asymmetry closely parallels the dynamics
of the spectral diffusion of the OH stretch peak, with a fast component that
is mostly associated with the intermolecular O-O stretch motion, and a longer
time dynamics that is associated with the collective HB network
restructuring.\cite{Rey2002}


\section{Electronic Signature of the Asymmetry: XA Spectroscopy}
\label{sec:electr-sign-asymm}

\begin{figure}
  \centering
  \includegraphics[width=1.0\columnwidth]{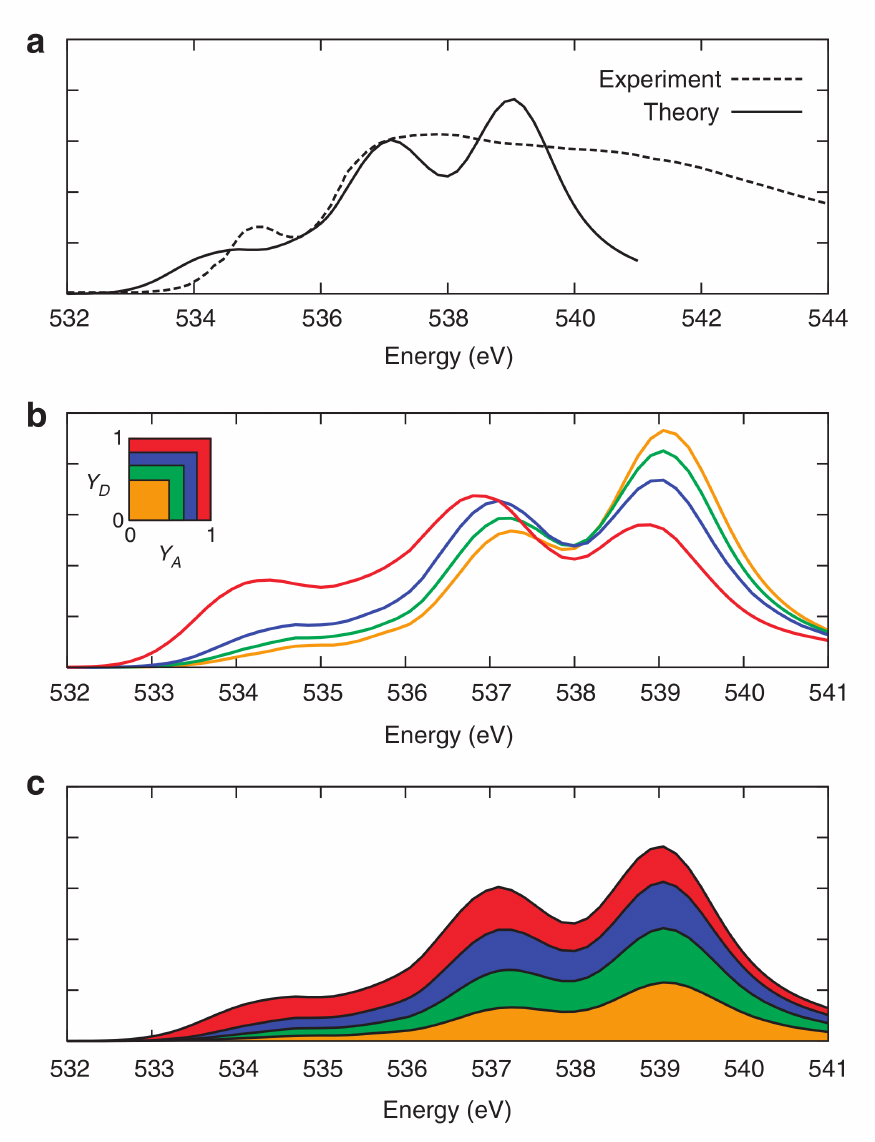}
  \caption{XA spectra of liquid water. (a) Comparison of the calculated and
    experimental\cite{Wernet2004} XA spectra. (b) Calculated XA spectra of the
    four groups of molecules separated according to the asymmetry of their
    donor ($\gamma_D$) and acceptor ($\gamma_A$) environments, as shown in the
    inset. (c) Contributions of the four groups into the total XA spectrum.
    The colour coding is shown in the inset above. Reproduced from Ref.~\citenum{Kuehne2013}.}
  \label{fig:xras}
\end{figure}

The time behaviour described above implies that the instantaneous asymmetry
can, in principle, be detected by X-ray spectroscopy, which has a temporal
resolution of several femtoseconds and is highly sensitive to perturbations in
the electronic structure of molecules.\cite{Wernet2004} To identify possible
relationships between the spectroscopic features and asymmetry, the X-ray
absorption (XA) spectrum of liquid water was calculated at the oxygen K-edge
using the half-core-hole transition potential formalism within all-electron
density functional theory.\cite{Iannuzzi2007, Iannuzzi2008, Muller2019} Although the employed computational approach
overestimates intensities in the post-edge part of the spectrum and
underestimates the pre-edge peak and overall spectral width, it provides an
accurate description of the core-level excitation processes and
semi-quantitatively reproduces the main features of the experimentally measured
spectra. The localized nature of the 1s core orbitals allows to disentangle the
spectral contributions from molecules with different asymmetry. To this end,
all molecules were separated into four groups according to the asymmetry of
their donor and acceptor environments, as shown in
\cref{fig:gamma_histogram}. As already mentioned, choosing boundaries at
$\gamma = \frac{1}{2},\frac{2}{3},\frac{5}{6}$ distributes all molecules into
four groups of approximately equal sizes (that is,
$25\pm2\%$). \Cref{fig:xras} shows four XA spectra obtained by averaging the
individual contributions of molecules in each group. It reveals that molecules
in the symmetric environment exhibits pronounced post-edge peaks, whereas molecules
with high asymmetry of their environment are characterized by the amplified
absorption in the pre-edge region. Furthermore, the relationship between the
asymmetry and absorption intensity is non-uniform: the pre-edge peak is
dramatically increased in the spectrum for the 25\% of molecules in the most
asymmetric group ($\gamma = \frac{5}{6}$), for which the strongest interaction is more than six
times stronger than the second strongest. As a consequence, the pre-edge feature of the
calculated XA is dominated by the contribution of molecules in the highly
asymmetric environments (\Cref{fig:xras}c). The pronounced pre-edge peak in
the experimentally measured XA spectrum of liquid water has been interpreted
as evidence for the so-called ``rings and chains'' structure, where $\sim 80\%$ of
molecules have two broken HBs.\cite{Wernet2004,Odelius2006} These
results however suggest that this feature of the XA spectrum can be explained by the
presence of a smaller fraction of water molecules with high instantaneous
asymmetry. Although the employed XA modelling methodology does not allow a precise
estimate of the size of this fraction, the result is consistent with
that of recent theoretical studies at an even higher level of theory, which
have demonstrated that the main features of the experimental XA spectra can be
reproduced in simulations based on conventional nearly tetrahedral
models.\cite{Chen2010,Kong2012b}

Thus, the investigation of the relation between local inhomogeneities in the HB
network and the XA spectrum revealed an interesting and important connection
between relatively small geometric perturbations in the HB network and
the large asymmetry in the electronic ground state, as well as the XA spectral
signatures of the core-excitation processes. This helps to reconcile the two
existing and seemingly opposite models of liquid water --- the traditional
symmetric and the more recently proposed asymmetric model.

\section{Vibrational Signature of the Asymmetry: Inhomogeneous broadening of
  the O-H stretch peak}

On the one hand, the structural picture obtained from XA spectroscopy essentially corresponds
to snapshots of the system that are frozen on the time scales of nuclear
motion. Raman and IR spectroscopies, on the other hand, directly
probe the resonance frequencies of the vibrational motions of the ionic cores
of the atoms. Moreover, the intrinsic time resolution of time-resolved IR
spectroscopy is about \SI{50}{\femto\second}.\cite{Hamm2011} It is also well
known that the strength of a HB is correlated with the frequency of the
covalent OH bond stretch motion, with stronger HBs being associated with a
red-shift in the frequency.\cite{Lawrence2002b,Lawrence2003} In fact, it is
mainly because of the disorder-induced inhomogeneities in HB strength that
the line width of the OH stretch peak in vibrational spectra of liquid
water are much broader than that in the gas-phase, \emph{i.e.} they are inhomogeneously broadened with some contribution from homogeneous broadening
as well.\cite{Perakis2016,Lawrence2002b,Bakker2010} Furthermore, there seems
not to be a substantial extent of motional narrowing and thus the total
linewidth is close to the inhomogeneous limit.\cite{Lawrence2002b} Taken these
features of the OH stretch peak together, one can immediately conclude
that the instantaneous asymmetry might exhibit some observable effects within
vibrational spectroscopy. In fact, it is tempting to say that vibrational
spectroscopy, and its lower frequency sibling, THz spectroscopy, are the
natural techniques to probe any structural or dynamic manifestations of the
local HB asymmetry.



\begin{figure}
  \centering \includegraphics[width=1.0\columnwidth]{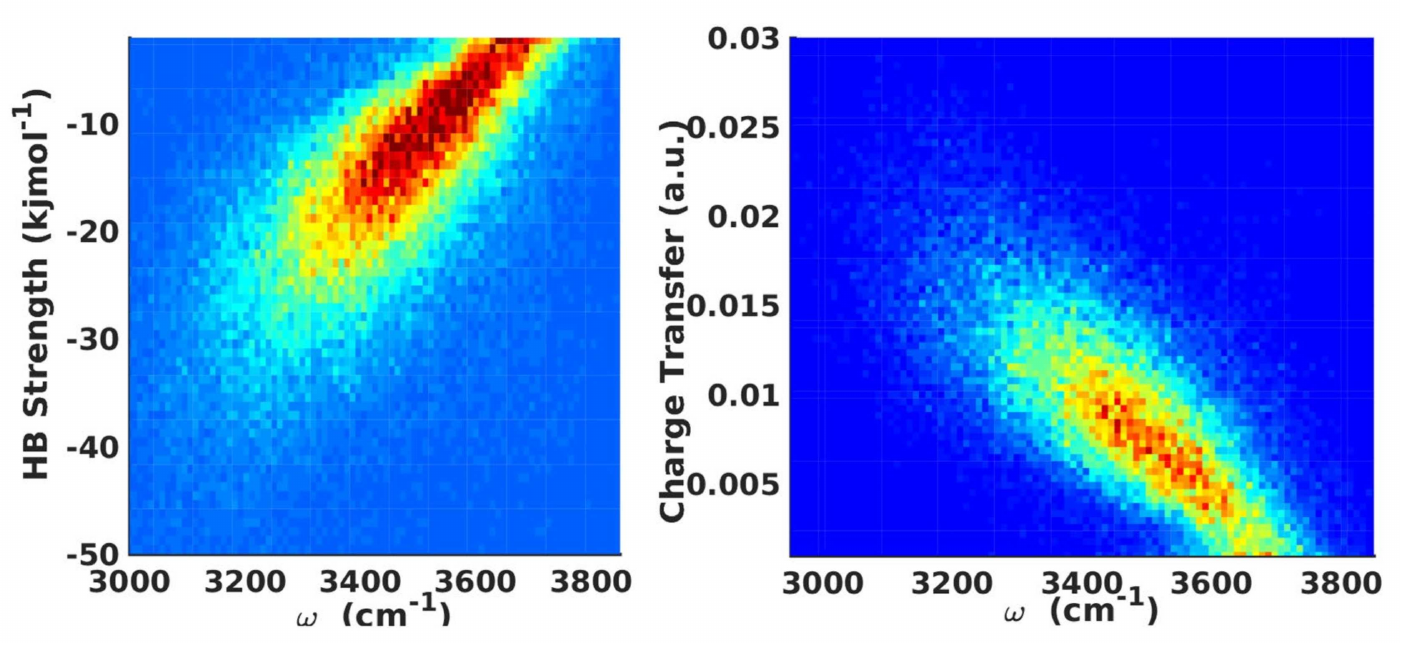}
  \caption{Equilibrium distributions of OH stretch frequency versus HB
    strength, as quantified by the $\Delta{E}_{CT}$ terms (left) and the associated charge-transfer terms ($\Delta{Q}_{CT}$ in units of elementary charge, right). The frequencies of the OH modes were calculated using a wavelet-based time-series
    analysis.\cite{Ojha2015,Ojha2018b,Ojha2018} Reproduced from Ref.~\citenum{Ojha2018} / CC BY.}
  \label{fig:deepak}
\end{figure}

As already alluded to above, ALMO-EDA offers a powerful
and consistent way to quantify the strength of HB interactions in liquid
water, and hence to \emph{quantitatively} establish the aforementioned
relationships between the structure and dynamics of the HB network and its
characteristics on the OH IR and Raman peaks, as obtained from
linear, non-linear, and time-resolved techniques.\cite{Bakker2010} Along these
lines of reasoning, ALMO-EDA was used to investigate the relationship between
the vibrational fluctuations of the OH stretching modes and the strength of
the HB, as quantified using the $\Delta{E}_{CT}$
(\cref{fig:deepak}).\cite{Ojha2018,Ojha2019} The instantaneous fluctuations in frequency of the OH modes were calculated using a wavelet-based time-series
analysis.\cite{Ojha2015,Ojha2018b}
The clear correlation in \cref{fig:deepak} indeed shows that $\Delta{E}_{CT}$,
taken as a measure of HB strength, linearly correlates with the frequency of
the covalent OH stretch vibration. A linear regression yields
\begin{equation}
  \Delta{E}_{CT} (\si{\kilo\joule\per\mole})= 0.0392 - 150.6\omega\,
  (\si{\per\centi\meter}), 
\end{equation}
which can be used to estimate the average charge-transfer-associated stabilization of a HB, corresponding to a given OH frequency of the electron-accepting water molecule. The root mean square error of the fit is \SI{7.27}{\kilo\joule\per\mole}. Such a large dispersion of the
HB strength in relation to the OH vibrational frequency (and vice versa) agrees well with previous findings regarding the relation between O-O HB length and
the OH frequency.\cite{Rey2002} It reaffirms our previous remark that the
relation between the HB strength and the local environment cannot be trivially
inferred from a single geometric or dynamical variable. It is also important
to note that for similar reasons, the dispersion of the observed data points
varies with the frequency/HB strength, stronger HBs exhibit a higher variation
in the associated OH frequencies.

\begin{figure}
  \centering
  \includegraphics[width=1.0\columnwidth]{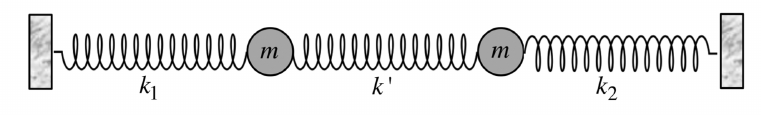}
  \caption{Coupled harmonic oscillators with mass $m$ and force constants
    $k_1$, $k_2$, and $k'$. Reprinted with permission from Ref.~\cite{Zhang2013}. Copyright 2013 American Chemical Society.}
  \label{fig:coupled_osc}
\end{figure}

\begin{figure}
  \centering
  \includegraphics[width=1.0\columnwidth]{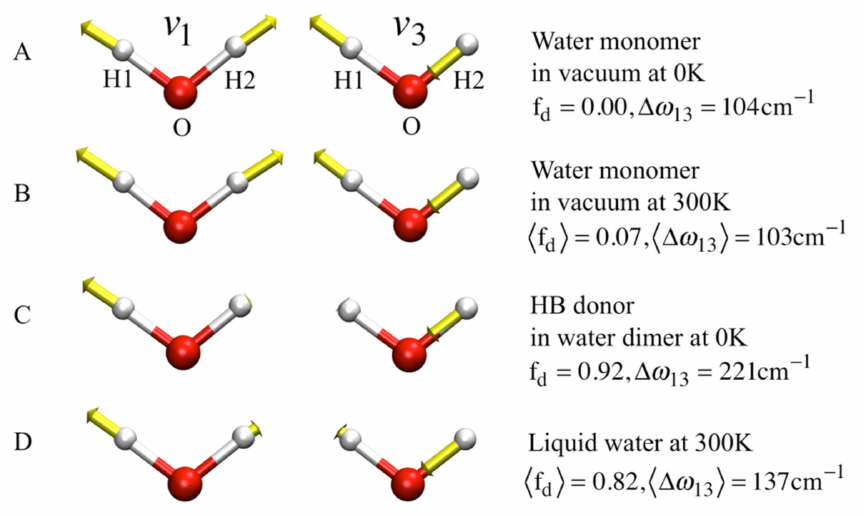}
  \caption{The symmetric ($\nu_1$) and asymmetric ($\nu_3$) stretching modes
    of a water molecule and their splitting frequency
    $\Delta{\omega}_{13} = \omega_3-\omega_1$ (A) in vacuum at \SI{0}{\kelvin}, (B)
    in vacuum at \SI{300}{\kelvin}, (C) of the H-Bond donor of a water dimer
    at \SI{0}{\kelvin}, and (D) in liquid water at \SI{300}{\kelvin}, respectively. See
    \cref{eq:fd} for the definition of $\textrm{f}_{\textrm{d}}$. Reprinted with permission from Ref.~\citenum{Zhang2013}. Copyright 2013 American Chemical Society.}
  \label{fig:nmodes}
\end{figure}

Most importantly, the linear trend in \cref{fig:deepak} suggests a mean to
unravel the consequences of asymmetry on the inhomogeneously-broadened OH
stretch peak. In order to see how this is possible, let us now describe the
two O-H stretch vibrations on a single water molecule as two coupled harmonic
oscillators which we denote as O-H1 and O-H2, with the same mass $m$ and restoring force constants $k_1$ and $k_2$ that are coupled by the intramolecular harmonic
coupling constant $k'$ (see \cref{fig:coupled_osc}).
If $k_1 = k_2$, as would be the case for water in the gas phase, or for a
symmetric HB environment in condensed phase, then the coupled normal modes
of the system $\nu_1= [1,1]$ and $\nu_3= [-1,1]$ are the familiar symmetric and asymmetric stretch modes. The energy splitting between the two eigenmodes is given by
$\Delta{\omega}_{13} = \omega_3-\omega_1 = k'/\sqrt{m(k_{1=3}+k)}$ ($k_{1=3}$ is used
to denote that $k_1=k_3$ in this case). We can see in this case that both
modes equally contribute to the two coupled normal modes, \emph{i.e.}  the two
O-H stretch vibrations are perfectly and symmetrically coupled, and the
hydrogen atoms move together with the same amplitude. Harmonic mode analysis
on a water monomer in vacuum using DFT gives a splitting
$\Delta{\omega}_{13} = \SI{104}{\per\centi\meter}$ at \SI{0}{\kelvin}
(\cref{fig:nmodes}A).\cite{Zhang2013} For comparison, CCSDT(T)
calculations at the complete basis set limit yield
$\Delta{\omega}_{13} = \SI{109.8}{\per\centi\meter}$.\cite{Kim1995}
However, when $k_1 \neq k_2$, the two normal modes are $\nu_{1,3} = [\delta \pm \sqrt{1+\delta^2},1]$, with a splitting
$\Delta{\omega}_{13} \propto (1+\delta^2)$ where
$\delta = (k_2-k_1)/2k'$.\cite{Zhang2013} The asymmetry in the restoring force
leads to a decoupling of the two O-H stretch vibrations, and the degree of
decoupling depends on the difference between $k_1$ and $k_2$. In fact, ab initio calculations reveal that the decoupling of the two O-H stretch modes in
the HB donor of the water dimer is almost complete, where each mode being
composed roughly of 85\% from one O-H stretch, with only a 15\% contribution
from the other O-H being mixed in.\cite{Kalescky2012} In this case normal mode
analysis shows, as expected, that the frequency splitting increases to
\SI{221}{\per\centi\meter} (\cref{fig:nmodes}C).

The simple model of two coupled harmonic oscillators shows that one consequence of
the local asymmetry in the HB environment is to decouple the two
intramolecular O-H bond vibrations, as well as to increase the frequency splitting
between the resultant symmetric and the asymmetric stretch modes. Indeed, it
becomes possible to quantify the HB asymmetry with the degree of
decoupling $\FD{}$ between the two O-H bond vibrations
\begin{equation}
  \FD{} = \sum_{k=\nu_1,\nu_3}\frac{|C_{KO-H1} - C_{KO-H2}|}{2(C_{KO-H1} +
    C_{KO-H2})}
  \label{eq:fd}
\end{equation}
where $C_{KO-H1}$ and $C_{KO-H2}$ are the contributions of the O-H1 and O-H2 bonds to the normal mode $k \in \{\nu_1,\nu_2\}$, respectively. When
$\textrm{f}_{\textrm{d}}$ is equal to 0, the two O-H bonds are completely
coupled in their motions. An increase in $\textrm{f}_\textrm{d}$, just like an
increase in the frequency splitting, is a signature of water molecules in
asymmetric HB environments and, according to the previous findings, should
be manifest in liquid water.
This line of reasoning has been tested with AIMD simulations of liquid
water.\cite{Zhang2013} To this end, a pre-requisite is to extract the
localized normal modes ($\nu_1$ and $\nu_3$) from the total vibrational
density of states of the bulk liquid, because the latter is the quantity that is directly accessible from MD simulations. One way to achieve this is by
requiring maximal localization of the power spectra of the two local modes in
the frequency domain. If we take the localization criterion as a minimization
of the quantity $\langle\omega^{2n}\rangle - \langle\omega^n\rangle$, where $n$ is an integer constant, then this leads to the requirement of minimizing
the following functional\cite{Martinez2006}:
\begin{eqnarray}
  \Omega^{(n)}=&\sum_k \left ( \frac{\beta}{2\pi} \int \textrm{d}\omega
                 |\omega|^{2n} P^{\dot{\nu}_k}(\omega) \right. \nonumber \\
               & - \left. \left ( \frac{\beta}{2\pi} \int \textrm{d}\omega
                 |\omega|^{n} P^{\dot{\nu}_k}(\omega)  \right )^2   \right )
                 \label{eq:martinez}
\end{eqnarray}
with respect to $\nu_k$, where $\beta = 1/k_BT$, and $P^{\dot{\nu}_k}$ is the
vibrational density of states of $\nu_{k}$. It can be shown that for $n=2$,
this method is equivalent to a normal mode analysis performed on the thermally averaged
Hessian matrix and generalized to anharmonic systems at finite temperatures,
and that the zero-temperature limit yields the usual normal modes.\cite{Martinez2006}

With this procedure, the two normal modes of each water molecule, $\nu_1$ and
$\nu_3$, were extracted from MD trajectories at finite temperatures. A
splitting of $\SI{103}{\per\centi\meter}$ was obtained from MD trajectories of
a single water molecule in vacuum at \SI{300}{\kelvin}
(\cref{fig:nmodes}B). The essentially identical results obtained at
\SI{0}{\kelvin} (using the familiar normal mode analysis) and
\SI{300}{\kelvin} (from MD trajectories) not only validate this approach, but
they further indicate that temperature and anharmonicity effects alone have
negligible influence on the mode coupling. The average value of the parameter
$\FD{}$ turns out to be 0.07, in clear contrast to the water dimer case with
$\langle\FD{}\rangle=0.92$, thus indicating an almost full decoupling between
the two O-H stretch vibrations of the HB donor molecule, as already pointed
out.
Finally, in the case of liquid water at \SI{300}{\kelvin},
$\langle\FD{}\rangle$ turns out to be 0.82, with an average splitting of
\SI{137}{\per\centi\meter} (\cref{fig:nmodes}D). In this case, the modes resembles those of the water dimer and, interestingly, also share similarities with the instantaneous normal modes of water molecules at water/vapor interfaces.\cite{Perry2003} Interestingly, it was found that for
$\FD{}\approx 1$ and $\Delta{\omega}_{13} > \SI{400}{\per\centi\meter}$, the
dipole moment of water molecules shifts to a lower value by $\SI{0.25}{D}$
with respect to the fully symmetric case in bulk water. This resembles
interfacial water molecules in water/vapor systems.\cite{Zhang2015}
By empirically fitting the data obtained from MD, it was also shown that the
relation between $\FD{}$ and the frequency splitting is closely approximated
by $\Delta{\omega_{13}}= C(1+\FD{}^2)$, where $C$ is an empirical fitting
parameter. This relation holds in the range $\FD=0$ until $\FD{} \approx 0.8$,
after which, the highly asymmetric configurations of water exhibit a broader
distribution in $\Delta{\omega}_{13}$ and a skew towards higher frequency
splitting. For the configurations, where a HB is totally broken,
$\Delta{\omega_{13}}$ increases up to more than \SI{400}{\per\centi\meter}.

\begin{figure}
  \centering
  \includegraphics[width=1.0\columnwidth]{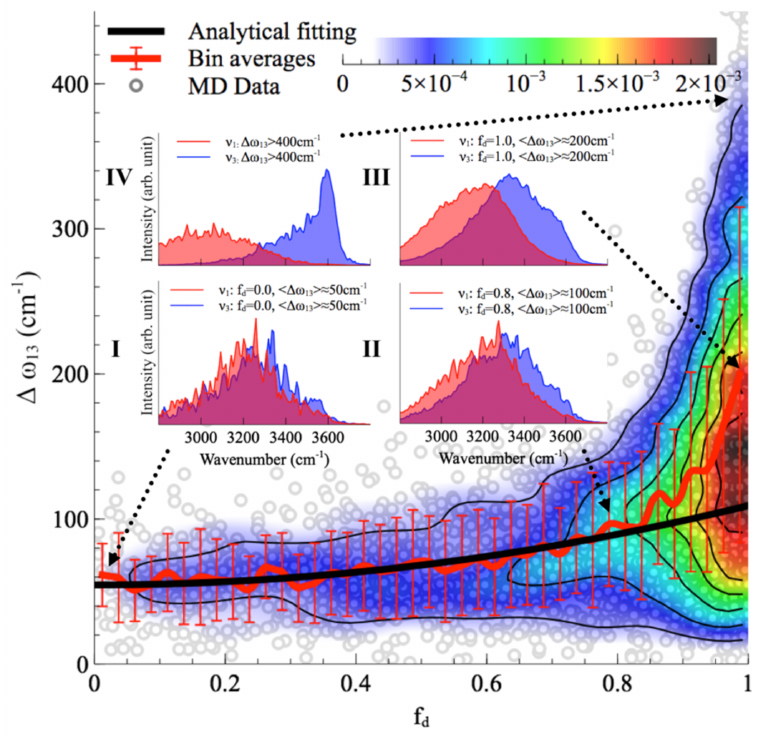}
  \caption{The normalized joint distribution of vibrational descriptors $\FD{}$ and $\Delta{\omega}_{13}$ for liquid water. $\langle \Delta{\omega}_{13} \rangle$ as a function of $\FD{}$ is also shown. Error bars denote the standard deviation of $\langle \Delta{\omega}_{13} \rangle$ for each value of $\FD{}$ with a bin size of 0.025. The continuous line has been obtained by fitting the data of $\langle \Delta{\omega}_{13} \rangle$ with the expression $\langle \Delta{\omega}_{13} \rangle \propto 1+\FD{}^2$. Insets I -- IV denote the vibrational spectra of the symmetric and asymmetric stretching modes $\nu_1$ and $\nu_3$ of the four representative regions in the $(\FD{},\Delta{\omega}_{13})$ space. The corresponding vibrational spectra of $\nu_1$ and $\nu_3$ are the averages of the decomposed vibrational density of states of configurations within each region. Reprinted with permission from Ref.~\citenum{Zhang2013}. Copyright 2013 American Chemical Society.}
  \label{fig:fd_domega}
\end{figure}

Although the asymmetry in HB strength seems to be the strongest factor in
determining the frequency splitting, and hence, the inhomogeneous broadening
of the O-H stretch peak in liquid water, other factors exhibit a role as
well. In a later investigation the role of the intramolecular coupling between
the two O-H stretch modes ($k'$) was scrutinized.\cite{Zhang2015} It was found
that $k'$ does modulate the frequency splitting, in a reverse fashion to
$\FD$, with the overall effect that the observed frequency splitting in liquid
water is less than the value that is predicted solely based on the effect of
$\FD{}$.
As a consequence, the instantaneous asymmetry of the local HB environment
around a water molecule in bulk water was shown to account to the observed
inhomogeneous broadening of the O-H stretch peak. While the role played by the
finite-temperature distribution of HB strength in the O-H stretch, inhomogeneous broadening in neither surprising, nor a novel finding (see
Ref. \citenum{Lawrence2003} for instance), instead, the new aspect here is that the
distribution of HB strength is associated with the peculiar feature of
mostly coming in pairs of strong-weak HBs, which are localized on the water
molecules. A feature that is most clearly seen on the right side of
\cref{fig:fd_domega} as $\FD{}$ approaches 1.


\section{Enhancement of Asymmetry Under External Electric Fields}

The study of water under electric fields is immensely interesting for several
reasons. First, water does exist under moderately to substantially strong
electric fields in a variety of natural settings, for example within
biological membrane channels, in the vicinity of electrodes and ions in
solution, and in cracks at crystal surfaces. The electric field intensity in
some of these cases can be of the order of
\SI{e9}{\volt\per\meter}.\cite{Rasaiah2008,Toney1994,Gavish1992} Another
reason of interest for studying water under electric fields, is that the
electric field-induced anisotropy can give rise to new interesting
features\cite{Evans1985,Skinner2012,Ho2013,Shafiei2019,Shafiei2019b,Baer2019} and can also enhance the
equilibrium structural/dynamical features in a liquid, facilitating their
study and hence providing new ways to understand the complexities of liquid
state kinetics and thermodynamics.\cite{Evans1982b} A good case 
for the latter is the long-established study of dielectric relaxation both
experimentally and theoretically,\cite{Coffey2004,Zasetsky2011} and for the
former, the discovery of non-vanishing rotational-translational
cross-correlations in water under electric fields.\cite{Evans1983c,Evans1985,Baer2019}
External electric fields, both static and fluctuating, are known to induce a
variety of field-induced anisotropies in liquid
water.\cite{Vegiri2004,Futera2017} In the context of HB asymmetry, the
significance of application of an external electric field lies in the
possibility of exploiting these field-induced anisotropies to enhance the
asymmetry, thus for instance to facilitate its experimental
investigation. Naturally, in this case, the origin of the --- possibly enhanced
--- anisotropy would not be the thermal fluctuations, but rather the external
field itself.

\begin{figure*}
  \centering
  \includegraphics[width=2.0\columnwidth]{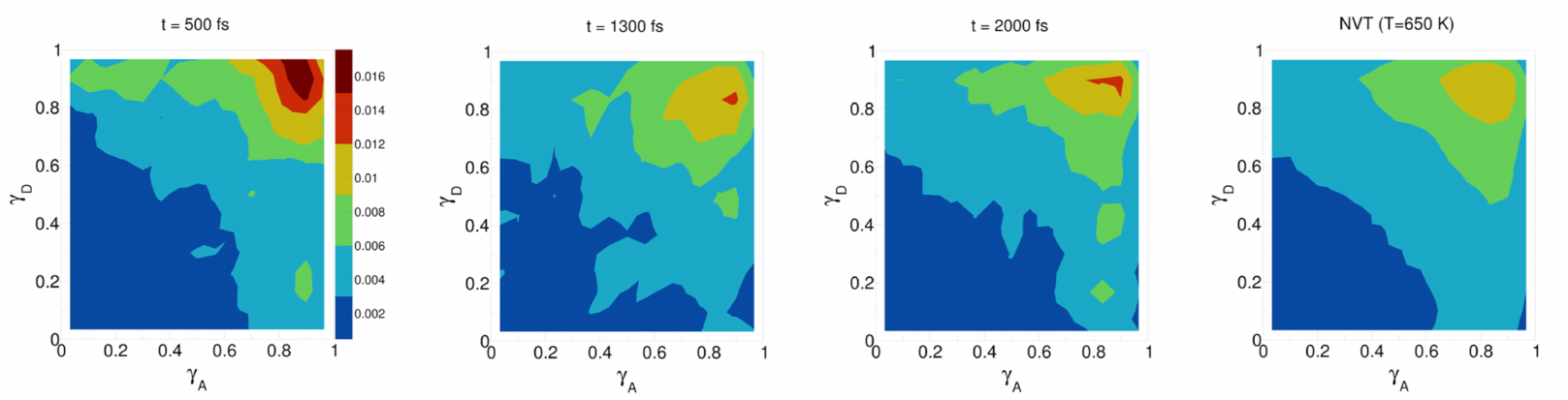}
  \caption{Progression of the joint distribution of the dimensionless
    asymmetry parameters $\gamma_A$ and $\gamma_D$ after an intense ultrafast
    rectangular electric field pulse. Time $t$ = 500 fs corresponds to the end
    of the pulse.}
  \label{fig:efield_asym}
\end{figure*}

The effect of an electric field on the local HB asymmetry in water was
investigated with AIMD using an intense electric field square pulse
with an intensity of \SI{4.3e9}{\volt\per\meter}.\cite{Elgabarty2019} The
pulse strength was chosen such that it induces an ultrafast re-orientation of
the water molecules on a sub-picosecond time scale. Under these conditions, it
was shown that within \SI{300}{\femto\second}, the water molecules reach a
steady-state average orientation angle of 37 degrees
($\langle cos(\theta) \rangle \approx 0.8$), where the angle is calculated
between the molecular bisector and the direction of the external field. This
ballistic re-orientation of the water molecules substantially increases the
temperature of the system up to \SI{650}{\kelvin}. Once the pulse is switched
off, this orientational anisotropy decays exponentially and vanishes within
\SI{750}{\femto\second}.
The influence of the pulse on the asymmetry of the HB network is shown in \Cref{fig:efield_asym}, which depicts the joint probability distribution of the asymmetry parameters $\gamma_A$ and $\gamma_D$ at various
times after the pulse. To distinguish the electric field induced effects from
effects that are only due to the high temperature of the system (in particular
the drop in the HB density), the joint probability distribution was compared
to that found in a field-free microcanonical trajectory simulated at an
average temperature of 650 K (right-most plot of
\Cref{fig:efield_asym}). We see in \Cref{fig:efield_asym} that immediately
following the pulse, the probability distribution has its peak at the top
right corner of the plot, where the molecules exhibit a high level of
asymmetry simultaneously in the two asymmetry parameters. The asymmetry
patterns in \cref{fig:efield_asym} are very distinct from the situation in liquid water under
ambient conditions, where the largest population of molecules exhibits high
asymmetry in one, but not in both asymmetry parameters. This is also
distinctively different from the field-free situation in hexagonal ice, where
the asymmetries in both parameters are uncorrelated.\cite{Kuehne2014} Comparison to field-free conditions shows that the electric field appreciably enhances the asymmetry. This enhanced asymmetry
then gradually decays once the field is switched off, so that after 1 ps, the
joint distribution has almost fully relaxed to the situation found in the
high-temperature trajectory.

The enhanced HB asymmetry under an electric field can be explained by the
field-induced anisotropy in HB strength. It was found that the pulse leads to
a strong anisotropy in the HB orientation along with the field-induced
molecular re-orientation. Moreover, the better a HB is aligned with the field,
the shorter it is. This arrangement makes charge-transfer through the HB more
favorable as it corresponds to a lowering in the electric potential along the
HB, and thus leads to an overall strengthening of the HB. These findings agree with the directional pair density distributions that were recently reported from simulations of water under electric fields.\cite{Shafiei2019,Shafiei2019b,Baer2019} After the pulse, the HB orientational anisotropy decays on the same time scale as the molecular one. Consequently, it was found that the strongest acceptor or donor
interaction is typically pointing along the field axis, while the weaker one
is more-or-less in the orthogonal plane. The molecules that are simultaneously
engaged in two HBs and exhibit a high degree of asymmetry ($\gamma > 0.8$),
are those simultaneously donating (or accepting) one HB in parallel and
another one in an orientation that is more-or-less orthogonal to the field.

\section{Water under a single THz Pulse: Influence of asymmetry on molecular
  re-orientation}

For a long time, one obstacle that has hindered a better understanding of water's HB network has been the absence of an experimental technique that directly probes this network. The pico- to sub-picosecond lifetimes of HBs\cite{Luzar2000} are too short for the NMR and dielectric spectroscopy time window and is only indirectly accessible by time-resolved IR spectroscopy.\cite{Shiraga2014} The recent technological feasibility of intense THz laser pulses has lately changed this state of affairs, opening new possibilities for the direct coherent excitation and control of the intermolecular and collective HB modes in water.\cite{Sajadi2017,Kampfrath2018,Baxter2011,Zalden2018}

\begin{figure}
  \centering \includegraphics[width=0.8\columnwidth]{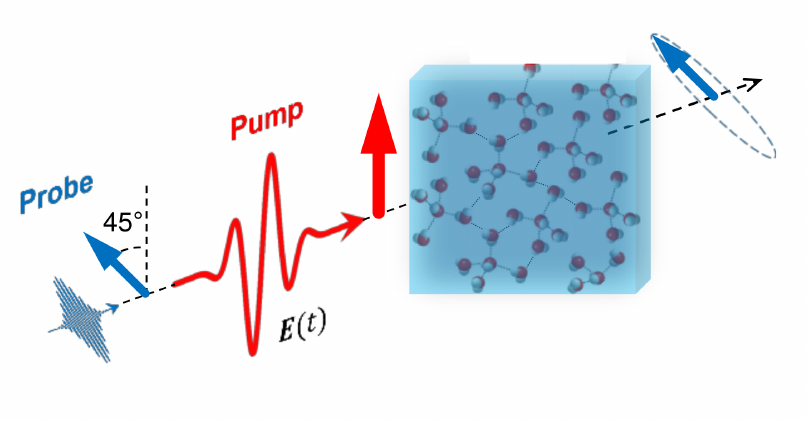}
  \caption{Dynamic THz Kerr effect. An intense THz pump pulse is used to induce optical birefringence in water, which is monitored by an optical probe pulse that becomes elliptically polarized upon traversing through the medium.\cite{Kampfrath2018a,Elgabarty2020}}
  \label{fig:tke}
\end{figure}

\begin{figure}
  \centering \includegraphics[width=1.0\columnwidth]{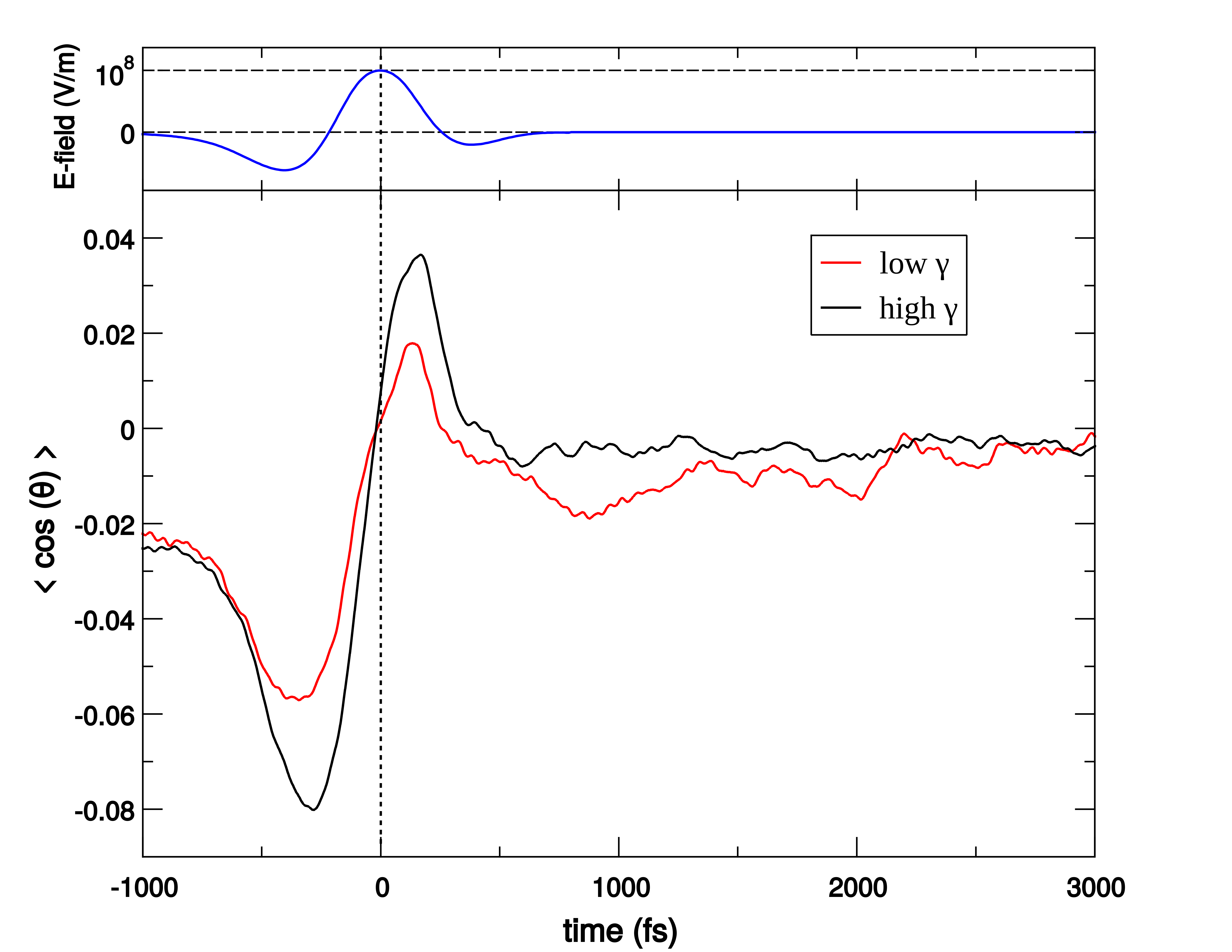}
  \caption{Dielectric alignment of water molecules with high (black) and low (red) asymmetry parameter under a single THz pulse. High-$\gamma$ is defined as $\gamma >0.8$, whereas low-$\gamma$ corresponds to $\gamma < 0.2$. The average angle between the molecular bisector and the field axis is denoted as $\cos{\theta}$. The vertical dotted line marks the reference point in time at which $\gamma$ was calculated and
    corresponds to the peak field in the THz pulse. The THz field profile is shown in the top panel.}
  \label{fig:thz}
\end{figure}

Leveraging these experimental advances, we have recently investigated the (sub)picosecond pathways of energy transfer from a THz pulse to water, using a novel THz experimental setup (\cref{fig:tke}) in combination with MD simulations.\cite{Elgabarty2020} So far, the physical nature of such fast relaxation processes within the HB network of water has been poorly understood and is heavily debated.\cite{Zasetsky2011} Our study has elucidated that the energy of a single THz pulse with a frequency of \SI{1}{\tera\hertz} couples mostly to the rotational dynamics of the water molecules, with 85\% of the energy going directly to rotational motions and only 15\% of the energy going into restricted molecular translational motion. The highly efficient ro-translational coupling in liquid water, coupled with the rigidity of the liberational potential (15 -- 20~\si{\tera\hertz}), then leads to an increase of the molecular translational kinetic energy of water molecules after the pulse, which lasts for $\sim\SI{1}{\pico\second}$ and decays on the same time scale as the observed THz Kerr effect. Interestingly, THz spectroscopy can also shed some light on the local asymmetry of the HB network. When we examined the relation between field-induced molecular re-orientation, and the asymmetry parameters $\gamma$, we found that the molecules with high asymmetry ($\gamma_{A/D} > 0.8$) are orientationally more labile than the molecules with low $\gamma$ ($\gamma_{A/D} < 0.2$). The average angle between the molecular bisectors and the field axis is depicted in \Cref{fig:thz} for high- and low-$\gamma$, where $\gamma$ was calculated at the time corresponding to the peak THz field intensity (\SI{e8}{\volt\per\meter}), denoted as $t=0$ on the x-axis. As expected from the lifetime of the asymmetry, the impact on the orientational dynamics can be observed for a few hundred femtoseconds before and after the point in time where $\gamma$ is calculated. We believe that focusing on HB strengths and asymmetries in these strengths, can be a more fruitful pathway than trying to identify geometric defects in the structure of water, and can help us understanding the relationship between orientational relaxation of a single molecule and the collective relaxation, a problem which remains at the frontier in studies of orientational relaxation.\cite{Bagchi2012,Zasetsky2011}



\section{Conclusions}

It is far from surprising that in the long-standing and ongoing debates regarding water
structure and dynamics, questions about the structure and dynamics of the HB
network are at the center of the stage. For example, in the invocation of
various two-state models of liquid water to explain thermodynamic or
spectroscopic observables (see for instance
Ref.~\citenum{Roentgen1892,Wernet2004,Russo2018,Gallo2016}), in the hypothesis
of a small population of mobile fast re-orienting water molecules to explain
dielectric relaxation,\cite{Buchner1999,Tielrooij2010,Zasetsky2011} and in various
hypotheses of local structural defects and their
propagation.\cite{Popov2016,Elton2017} In all these cases, the core
assumptions are fundamentally about the structure and dynamics of the HB
network on different time scales, its ability to spatially and temporally
sustain local defects, the life times of these defects, and their spatial
propagation through the network. This is to say that, as a means towards
fruitful progress, a possible falsification of any of these hypotheses must be
ultimately based on a better understanding of the dynamics of the HB
network. In this regard, it is unfortunate that in many cases the
indiscriminate use of descriptive words like ``structure'' and
``heterogeneity'' can lead to much confusion if not accompanied by clear
operational definitions and a context under which these definitions are
valid. The literature on water structure and dynamics is full of descriptions
such as ``enhanced structure'', ``structure formation/breaking'', and
``dynamic polymer'', just to name a few. Such explanations can be useful and
insightful, but they also have the potential to quickly become mere issues of
semantics and nothing more. We agree with previous criticisms that the
indiscriminate use of descriptive words in the water literature has only lead
to much confusion.\cite{Qvist2008}


In this article, we have discussed an aspect of the HB network that has been
recently unraveled using a combination of AIMD
together with ALMO-EDA.\cite{Khaliullin2013} We have shown that in equilibrium
liquid water, at any instance of time, a large population of water molecules
have a high disparity between the strengths of the two HBs they are donating
or the two HBs they are accepting, an aspect which we refer to as ``local
asymmetry in the HB network''. Here, the strength of the HB is quantified by
the amount of charge-transfer from the HB-acceptor to the HB-donor and the
associated energy lowering ($\Delta{E}_{CT}$), both quantities are obtained from
ALMO-EDA. We have also shown that the extent of asymmetry can be quantified
using the dimensionless asymmetry parameters $\gamma_A$ (electron acceptor
interaction) and $\gamma_D$ (electron donor interactions), where $\gamma$ runs
from zero (fully symmetric HB environment with two equally strong HB
interactions) to one (one HB is totally broken). Thus, we provide operationally defined
metrics that can be consistently employed on simulated water trajectories to
quantify the heterogeneity in the HB network and the dynamics of this
heterogeneity.

The idea that there is a broad distribution of HB strengths in liquid water is
neither novel nor surprising, and indeed our quantitative metrics of HB
strength confirms this picture, but furthermore, the local asymmetry presents a
stronger statement. It is neither necessary nor self-evident that a broad
distribution of HB strengths implies a significant population of water
molecules with a strong asymmetry in both asymmetry parameters simultaneously
(75\% of molecules with $\gamma_A$ and/or $\gamma_D$ greater than
0.5). This is particularly clear when comparing the picture in liquid water to
that in hexagonal ice. But again, it is important to emphasize here that the
picture we find is that of a continuous distribution of HB strengths rather
than any kind of two-state picture. When we discuss the behavior of high- versus low-$\gamma$ water molecules, it is meant to emphasize how the extremes of the
distribution are behaving, an aspect which is particularly important given the
significant abundance of molecules with high $\gamma$, but this is not to
imply that water is a mixture of two kinds of molecules or HBs. The
comparison with ice is once again illuminating in that it further corroborates
that the high asymmetry in water is a consequence of an interplay of thermal
disorder, the high connectivity and cooperativity of the HB network, and the
diffusive aspect of HB dynamics.

Regarding the traditional tetrahedral structure of liquid water, it is trivial
to realize that in a liquid system, what might be disordered on some short
time scale, becomes ordered when viewed over longer time scales, possibly just
as it is no more --- or less --- profound to realize that a liquid below some
relaxation time can be viewed as a solid glass and can similarly support
transverse and longitudinal phonon
modes.\cite{Frenkel1955,Trachenko2016,Elton2016} Keeping in mind the dynamics
of the asymmetry and how it decays, the strength of donor–acceptor
interactions we find, suggest that each molecule in liquid water at
ambient conditions forms, on average, two donor and two acceptor bonds. It is
because of the very strong dependence of the HB energy on the local geometry
(in particular the exponential dependence on the HB length) that even small thermal
distortions in the tetrahedral HB network induce a significant
asymmetry in the strength of the contacts causing one of the two donor
(acceptor) interactions to become, at any instance of time, substantially
stronger than the other. Thus, the instantaneous structure of water is
strongly asymmetric only according to the electronic criteria, not the
geometric one. Overall, the picture provided by ALMO-EDA does not warrant any
departure from the traditional tetrahedral structure.
With respect to the dynamics of the asymmetry, we have shown that it decays on a time scale of several hundred femtoseconds. Intermolecular vibrations (O-O
stretch) and librations of OH groups of HBs are primarily
responsible for the relaxation of the instantaneous asymmetry. The time scales, which we find, closely match those obtained from studying the time-dependence of
the OH spectral diffusion. The long-time non-exponential tail of the
relaxation seems to be related to the non-exponential behavior of HB kinetics,
which can be traced back to the translational diffusional aspect of the
kinetics.\cite{Luzar1996}

So what are the consequences of this asymmetry on the spectroscopic
observables of liquid water? The challenge in figuring out the answer is that the causal relation
between the HB energy (and its asymmetry) on the one hand, and the structure
and dynamics of water (\emph{e.g.} order parameters, power spectra, HB lengths and angles and their
distributions), on the other hand, is very far from trivial. We
know for instance that the removal of some of the HB partners in liquid water
has the potential to slow down water rotations, as it breaks cooperativity,
but the removal of all HBs, possibly by dispersion in non-polar solvents, can
greatly speed up the rotation of water molecules.\cite{Qvist2008,Halle2004,Kuehne2011,Kessler2015} So what happens when a water molecule is simultaneously donating a strong and a weak HB? This is somewhat like a water molecule that is tumbling on one strong and one weak leg. It turns out that in XA spectroscopy, this results in a population of water
molecules that gives an amplified response in the pre-edge peaks. In case of
the vibrational O-H stretch peak, it turns out that the ensuing decoupling
between the symmetric and antisymmetric normal modes can largely explain the
observed inhomogeneous broadening, whereas in pulsed THz spectroscopy, it turns
out that the molecules with high asymmetry are orientationally more ``labile''
on a time scale close to half a picosecond. Thus, we see that the manifestations are
present on different time scales spanning several orders of magnitude. Hence, trying to answer the question posed in the beginning of this paragraph, this can only be considered as work in progress. There are many more properties of water where asymmetry can play an important role, which remains to be investigated. Examples of the latter are attempts to explain the nature of energy dissipation processes,\cite{Elgabarty2020} and more generally, of fast relaxation processes in water, aspects that are still poorly understood and intensely debated.\cite{Buchner1999,Tielrooij2010,Zasetsky2011} Recent findings have also pointed towards an important role of intermolecular modes in vibrational energy relaxation, a role particularly played by strong HBs.\cite{Ramasesha2013} It would also be very interesting to see how asymmetry might dictate certain energy dissipation pathways in this case.

We owe much of what we know about liquid water to interpretations of experimental findings that were founded on force field MD simulations.\cite{Auer2007,Auer2008,Torii2006,Rey2009,Rey2015,Yagasaki2013,Medders2015a,Hamm2017b} It is an intriguing question whether any of the existing interpretations would be significantly modified if a force field that captures the asymmetry aspect were
employed. Recently, \citeauthor{Naserifar2019}\cite{Naserifar2019} have
published a work, where they describe the results of MD simulations of water
using a new force field parametrized using a combination of DFT and CCSD(T)
benchmark calculations. Interestingly, they find that each water molecule on
average has two strong and two weak HBs (based on a distance
criterion), and that the relaxation time between a strong and a week hydrogen
bond is $\sim\SI{100}{\femto\second}$. Despite of our agreement with the criticism that has been raised against this work\cite{HeadGordon2019}, these finding are intriguing and motivate further careful investigation and comparison with previous experimental and theoretical work.

\section*{Conflicts of interest}
There are no conflicts to declare.

\section*{Acknowledgements}
H.E. would like to acknowledge funding from the Deutsche Forschungsgemeinschaft --- DFG, for his temporary position as a principal investigator (Eigene Stelle). T.D.K. acknowledges funding from the European Research Council (ERC) under the European Union’s Horizon 2020 research and innovation program (grant agreement No 716142). The generous allocation of supercomputer time by the Paderborn Center for Parallel Computing (PC2) is kindly acknowledged.

\bibliography{All}
\bibliographystyle{rsc}

\end{document}